# MODEL OF INFORMATION SYSTEM TOWARDS HARMONIZED INDUSTRY AND COMPUTER SCIENCE

**A PROJECT WORK SUBMITTED IN PARTIAL FUFILMENT OF THE REQUIREMENT**

**FOR THE AWARD OF BACHELOR OF SCIENCE**

**B.Sc (HONS) DEGREE IN COMPUTER SCIENCE**

**SUBMITTED BY**

| | |
|---|---|
| **EDAFETANURE-IBEH FAITH** | **14/2841** |
| **EVAH PATRICK TAMARAUEFIYE** | **14/1590** |
| **MARK UWUORUYA UYI** | **14/1930** |

**SUPERVISED BY**

**DR ADEBAYO A.O**

**TO THE DEPARTMENT OF COMPUTER SCIENCE SCHOOL OF COMPUTING AND ENGINEERING**

**BABCOCK UNIVERSITY, ILISHAN REMO**

**OGUN STATE NIGERIA**

**MARCH, 2018**

## DEDICATION

This project is dedicated to God Almighty

## ACKNOWLEDGEMENT

Our most sincere gratitude goes out to Dr. Adebayo, our supervisor, for guiding and directing us through the course of working through and completing this project. Our gratitude also goes to our family for trust and support, emotionally and materially, during the course of this project. We would also like to show gratitude to staff in person of Prof. Adekunle, the H.O.D of our department and Dr. Alao, the project coordinator for their support and guidance during the course of this study. Also, we won’t forget extending our gratitude to the staff of the department for the knowledge passed and the guidance for our time spent in the institution.

## Contents

## LIST OF TABLES

## LIST OF FIGURES

## ABSTRACT

The aim of attending an educational institution is learning, which in turn is sought after for the reason of independence of thoughts, ideologies as well as physical and material independence. This physical and material independence is gotten from working in the industry, that is, being a part of the independent working population of the country. There needs to be a way by which students upon graduation can easily adapt to the real world with necessary skills and knowledge required. This problem has been a challenge in some computer science departments, which after effects known after the student begins to work in an industry. The objectives of this project include: Designing a web based chat application for the industry and computer science department, Develop a web based chat application for the industry and computer science and Evaluate the web based chat application for the industry and computer science department.

Waterfall system development lifecycle is used in establishing a system project plan, because it gives an overall list of processes and sub-processes required in developing a system. The descriptive research method applied in this project is documentary analysis of previous articles.

The result of the project is the design, software a web-based chat application that aids communication between the industry and the computer science department and the evaluation of the system. The application is able to store this information which can be decided to be used later.

Awareness of the software to companies and universities, implementation of the suggestions made by the industry in the computer science curriculum, use of this software in universities across Nigeria and use of this not just in the computer science field but in other field of study

# CHAPTER ONE

# INTRODUCTION

## 1.1 Background to the Study

The world today is quickly developing regarding the improvement of science and innovation. Almost all companies have an Information Communication Technology (ICT) department, and this creates a demand for better computer science graduates that are equipped with the required skills the industries need. However, industries often criticize the university curriculum, noting that it does not really equip students with the skills that will make them relevant in the industry (Ayofe & Ajetola, 2009).

An information system is assembling of data that a business uses, to make decision and also communicate with technology furtherance of business processes (Kroenke David, 2015). The old way of storing information on hard copy is getting obsolete, people wanted a simple and more effective way of storing information which is computer based information system. It has evolved overtime to be the backbone of all businesses today and not just businesses even our personal information. Some major companies, are built entirely around information system. Examples include Google, Facebook, Amazon, and the government who use information system to provide services for its citizens (Vladimir Zwass, 2011).

The level at which individuals use information system has affected the natural order of things. Traditional things like shopping, socializing, banking are now done with computers or phones in the comfort of homes. Information system helped dissemination of information to be much easier and also faster. An information system is defined as a collection of information that helps organize and analyze data. So, the purpose of an information system is to turn raw data into useful information that can be used for decision making in an organization (Paul Zandbergen, 2015).

This research work intends to create a method of bridging this gap between the industry and the computer science department of Babcock University. This research work will focus on a web platform that will achieve this purpose.

It is recommended that universities review their programs once in five years using current National Universities Commission (NUC) quality assurance benchmark statements. The review should incorporate the opinions of relevant stakeholders such as students, staff, employers and policy makers, among others (NUC, 2017).

**1.2 Statement of the Problem**

The model is aimed at solving the problem of lack of communication between the industry and the computer science department. The computer science department needs to know the requirement or need of an industry so as to be able to align these needs to the curriculum of the department. There needs to be a way by which students upon graduation can easily adapt to the real world with necessary skills and knowledge required. The school curriculum are mostly based on the old way practices and outdated things with little or no learning of the advances of the present day (Ingvi Hrannar, 2014). The school teaching methods are mostly theoretical and also little or no practical experiences (Ayofe & Ajetola, 2009). This problem has been a challenge in some computer science departments, which after effects known after the student begins to work in an industry.

## 1.3 Significance of Study

The significance of designing an information system for a harmonized industry and computer science cannot be over emphasized. Some of which is that it will help to improve the communication between the industry and computer science department. It will help in the reevaluation of the school's curriculum to align with that of the present day real world society.

It will also help in the transformation of the undergraduates from the mainly theoretical environment to the practical real world environment easier. Equip students with in-demand skills.

## 1.4 Research aim and Objectives

The goal of this project is to build a web-based application for computer science department to be able to evaluate their curriculum with the suggestions of the industry.

The objectives are to:

- Design a web-based chat application for the industry and computer science department
- Develop the web-based chat application for the industry and computer science department
- Evaluate the web-based chat application for the industry and computer science department

## 1.5 Methodology

Waterfall Model is the software development Life Cycle that would be used for the proposed system. The name of the software is the Industry Curriculum. The waterfall model would help in building the system faster and aid in proper documentation.

### 1.6 Scope of Study

The research project is carried out specifically to create an online web based platform to aid in communication between the industry and computer science.

### 1.7 Organization of Subsequent Chapters

This is the progression of the chapters from chapter one to five. It is explained as follows:

**Chapter One** – This chapter gives an introduction to the proposed project.

**Chapter two** – This chapter is the literature review. In this chapter, previous works and articles would be reviewed to better understand the problem and project.

**Chapter three -** This chapter is the system methodology. It gives procedures and methods in which the problem would be tackled. It also shows the design of the software.

**Chapter four-** This chapter is the system implementation. It shows the result of the software with screenshots and also a guide to how the software would be used.

**Chapter five –** This chapter is summary, conclusion and Recommendations. It gives a brief summary of the project, limitations faced and possible ways in which the software can be improved.

# CHAPTER TWO

# LITERAURE REVIEW

## 2.0 Introduction

This chapter reviews articles, journals, publication and recently performed research projects related to the current project. With the sole aim better understanding of the project topic to successfully executing the research.

### 2.0.1 Definition of Data

Data is facts that has been changed into a form that it can be processed effectively (Margaret Rouse, 2017). Information becomes information when it has been processed from the form of data.

### 2.0.2 Definition of Information

Information is information in the context of the person receiving it. When information is inputted into a computer, it said to be data but after it has been processed and probably printed, it can then said to be information again (Margaret Rouse, 2017).

## 2.1 Definition of System

A system could be an assortment of components or parts that area unit organized for a standard purpose. The word typically describes the organization or arrange itself and typically describes the components within the system (Margaret Rouse, 2017).

### 2.1.1 Definition of Information System

Information System typically is said because the backbone of the many businesses as a result of it handles the flow of knowledge and maintenance that supports businesses, with data of a company's day to day activities, their employees and properties (Margaret Rouse, 2017). Information is gotten after data has been processed. Data in its raw form may not be organized or easily understood but after processing, it can be easily understood by humans.

An Information System can be characterized in fact as an arrangement of correlative components that gather (or recover), process, store and allocate information to help with decision making and control in the industry. A system that combines, stores, processes, and convey data necessary for the industry or general public, in such a approach that the knowledge is available and advantageous to those that might want to utilize it, joint with employers, employees, purchasers and voters (Buckingham et al ,1987b). A system could act as a social system, which can or might not involve the utilization of pc systems. Also, additionally to promoting decision-making, data systems facilitate employees and managers to research advanced issues, to build new merchandise and to combine the assorted modules and departments. Additionally, there is better coordination and improved straight forwardness inside the organization because failures in communication and transmission have been reduced.

### 2.1.2 Types of Information System

The information system is broken down into six major types. The types of information system includes:

- Transaction Processing Systems (TPS): serve the operational level of an organization.t
- Knowledge work systems (KWS)
- Office automation systems (OAS) to serve the knowledge level of an organization.
- Decision-support systems (DSS)
- Management information systems (MIS) serve the management level of the organization.
- Executive support systems (ESS) serve the strategic level of an organization. (Kevin Cress, 2009)

### 2.1.3 Components of Information System

Software, Hardware, data, process and people are the five components of information system.

#### 2.1.3.1 Hardware

Hardware component is the component of the information system that can be touched. It can be said to be the physical components of the information system. Examples of hardware include keyboards, printers, computer and others (David Bourgeois, 2014)

#### 2.1.3.2 Software

Software directs the hardware on what to do by administering sets of instructions. Software unlike hardware cannot be felt .Among all the different divisions of software, there are two main divisions which is system software and application software. The system software makes the hardware usable, and application software, which helps in interaction. Examples of systems software include Apple's Mac OS, Microsoft Windows and Google's Android. Examples of application software are Microsoft Word and Microsoft PowerPoint. (David Bourgeois, 2014)

#### 2.1.3.3 Data

Data is the third element. You will be able to think of data as a group of raw facts. An example is a city, a town, and your address all items of knowledge. Data is similar to programs because they are both tangible .Data alone might be said not to be useful but when it has been processed and organized together into a database, data becomes very useful. Data is used in different field of study. An example is data mining. The amount of data in the world is growing at a very rapid and with data mining, the useful data are aggregated and organized. This helps businesses in decision making and helps predict future trends.

Apart from the elements of software, hardware, and data, that has been pondered upon the core technology of information systems, it has been argued that another alternative part ought to be added which is communication. The information system will exist even without the means of communication. The premier sets of personal computers which were stand-alone computers were incapable of accessing the internet. Nonetheless, in present day connected world, it is not common to see a laptop that does not connect to a network or with another device. Technically, hardware and software is what the networking communication part is formed, Nonetheless, it's such a core feature today's data systems that it's become its own class. (David Bourgeois, 2014)

### 2.1.3.4 People

In a case where information systems is been talked about, it's straightforward to urge targeted on the technology elements and forget that we have a tendency to should look on the far side these tools to totally perceive however they integrate into a company, attention on the people concerned in info systems is that the next step. From the front-line help-desk employees, to systems analysts, to programmers, all the high to the chief information officer (CIO), the people attached to information systems square measure a necessary component that has to not be unnoticed. (David Bourgeois, 2014)

### 2.1.3.5 Process

Process is the last component of information system. A process can be said to be a sequence of processes stipulated to acquire a desired result or output. Information systems have become more integrated with structure processes, transferal of a lot of productivity and higher management to those processes. However, merely automating activities making use of technology is not enough. Businesses require to effectively utilize information systems do more. Making use of technology to manage and improve processes, each among a corporation and outwardly with suppliers and customers, is the final goal. Technology all need to do with the continued improvement of those business procedures and therefore the integration of technology with them. Businesses hoping to realize a plus over their competitors square measure extremely targeted on this element of information systems. (David Bourgeois, 2014)

## 2.2 Related Articles

The related articles will be treated under three subject topics to better understand the problems and suggestions made by previous research. They include description of gaps, relevance of curriculum and bridging the gaps.

### 2.2.1 Description of Gaps

The primary reason for this gap is that existing computer science department syllabus is not aligned with that of the IT industry.(Zulfiqar, Jawaid, Javed, Sana, Irshad & Ghulam, 2017)

Apart from ability devolution that occurs after some time, there are alternative things that should be taken into consideration that causes this problem. A noteworthy issue is the dynamic trends of working in organizations. The current patterns inside the universe of labor like economic process, exploitation, freeing, allocated job, preparation and outsourcing have prompted checked changes in trade architecture. There has been new innovative findings that have offered ascension to new ventures and new way to organize work. There are new variations of work architectures which are multifaceted, flexible, and all-purpose and that energize constant acquisition of knowledge are getting wellsprings upper hand in organizations. Universal rivalry for occupations and staff has combined acute, bringing about the universal search for advanced innovation driven staff. Furthermore, businesses worldwide are preparing themselves to confront against the present day inadequacy to provide the necessary skills to the graduates which are used in the real world industry. In previous times, the overall information based economy relates regularly developing exceptional abilities, innovativeness, and capability of hands, pioneers of businesses articulate an augmenting hole between the ability their industries must be constrained to develop and furthermore the abilities of their workers. Business pioneers face the problem of finding the right contender to fill a developing rundown of empty positions. Investigation demonstrates that the movements in men statistics affect the accouterment of labor to occupy professional jobs. Amusingly, ability problem result from innovative improvement. Consequently, as a general rule, industries can constantly confront few types of skill gaps constantly if the college information does not alter itself into the innovative economy. Absence of right aptitudes inside the college students, absence of designing, absence of planning, perplexity, poor equipment for IT skills advancement and wasteful use of rare assets has greatly added to put our nation in an extremely problematic activity shortfall. The information innovative tutoring offices square measure couple of unorganized and unlooked among prominent industries. Prior to the slump of the world, the roles paucity was then cumbersome and encumbering. The case has right now turn out to be even more over essential. A litigant during a present examination apportioned endless supply of showing laborers and body troubles in change the college programs information for information technology instruction.

Absence of technical specialized understanding, expensive information technology devices, expensive upkeep and substitution of devices, are some of the chief obstructions. Another significant downside has been the universities ‘powerlessness to remain educated brisk dynamical improvements in exchange and innovation. It totally was built up before that specialty exists between subjects educated and furthermore the courses familiar tutor these courses, and the furthermore the instructional exercise necessities at teaching method foundation.

The Modern application is dynamic whereas the computer science curriculum is static. Absence of equipment to tutor the students and also the teachers on the current innovation. Lukewarm point of teachers to yield themselves for training and workshops which can expedite them to the latest developments in information technology. Need given to investigation works by the teachers rather than seminars and workshops which can expedite them to awareness of the latest innovation in information technology. Apart from this problem is deficiency of the academic syllabus which is planned and obvious respects for significance in use of the industries.(Ayofe & Ajetola,2009)

The need for graduates who are prepared for employment and skilled in teamwork has been widely advocated over the last decade.(Elisabeth & Mike,2000)

**2.2.2 Relevance of Curriculum**

Adjusting the syllabus to speedily dynamical desires of employers so the marketplace is therefore terribly imperative..(Ayofe & Ajetola,2009)

Results from the survey indicated that this subject offerings among the Computer department are indeed relevant to the requirements of business and therefore the geographic point but there are prospects or topics which require insertion or higher coverage. The study conjointly counseled usual surveys to determine connectedness of syllabus to desires of the business..(Ioana et al.,2015)

Computer science students typically are exposed to the breadth and depth of software system topics through a range of software system and programming courses. This typically isn't the case once it involves hardware topics. we have a tendency to powerfully feel that a balanced treatment of hardware and software system ideas can higher prepare our students to the subtle geographic point..(Krishnaprasad, 2002).

**2.2.3 Bridging the Gap**

The response to the present question clearly depends on presenting the researchers about the top-notch psychological feature abilities that area unit fundamental and required by companies. The sub topics discussed below are a couple of solutions that are known to produce logical outcomes:

• Learning IT Skills program: It is often common usually given either embedded in courses taught or as entire programs. Information technology skills are a unit strengthened, independent studying, long studying, investigative skills, management of time ability, critical reasoning ability and lots more. These parts are believed to be within reach after they have been inputted into the college syllabus rather than taking it has an entire courses. To overcome the issues or potentially improving the educational system, few nations have experienced and self-tended to this problem by acquainting a capable innovative part to the college information. It generally appears in a lot of different forms; common between them is giving students subjects in information technology, job edge and job ethic, which is accompanied by a chronological placement among industrial and industrial organizations, anywhere essential proficiency in actual job setting. Winning programs are implemented in nations like Japan, United States of America, Canada and Great Britain. The presence of big industrial sectors are the reason why such programs are successes , in which the industries partner with faculties. Different nations have selected to determine tutoring centers, which have seminars and workshops that provide students actual job proficiency. This tutoring centers unit usually sponsored and handled by the private sectors whom pay expenses for their students to make use of the facility. Some of the success of this kind of program is the Chicago school-to-Work Program and the BOCES program in big apple state.

• IT: The IT program guarantees that the coed has satisfactory data about information technology and possesses the capabilities expected to utilize it in the work. Two necessary part of IT that will be necessary for students who desire to join the work force is data and skills set. The syllabus or tutoring guidebooks that are implemented by several organizations are developed either completely by the International Journal of Engineering Science and Knowledge Security (IJCSIS) or by commissions like International Labor Organization, the National Board for Technical Education for the Polytechnics relevant coordinative commissions like the National

Universities Commission just in case of the schools, the National Board for Technical Education for the Polytechnics, and National Universities Commission. Satisfy it to specify that the fundamental focus of the academic structure is to deliver the necessary skills required by the general population and subsequently the sorted out private sector.

• Government including the private sector should ought to make course of action for talented students of tertiary institutions to hold up short-run reasonable tutoring inside their selected job by means of a scheme which is implemented in universities usually at three hundred or four hundred level, it is the Student Industrial Work Experience Scheme (SIWES) to support students in getting experience of the real world industry. An up rise in a of talented developers in information and Communication Technology, united nations agency would be put amongst institutional and individual programmers. The rise of this bunch is in light of advancements inside the ICT exchange. The government should make special endeavors towards handling activities inside the information technology sector to keep away from abeyant undesirable methods that may deface the field.

Modifying the syllabus dynamical desires of managers and work force is hence discouragingly obligatory. (Ayofe & Ajetola, 2009)

The Computing Curricula 2005 Task Force has thus sought-after to scale back the desired level of coverage in most areas to create area for brand spanking new content.(Ioana et al., 2015)

Practical tutoring inside their selected career job by a scheme known as the Student Industrial Work Experience Scheme (SIWES) which will help to promote the information in the sector.(Ayofe & Ajetola, 2009)

## 2.3 BMAS

Benchmark Minimum Academic Standard (BMAS) is meant to ensure quality assurance on university programs of Nigeria. The Computer Science department is the subject matter in this project.

### 2.3.1 Computer Science Course Outline

The computer science department course outline because of the Benchmark minimum academic standard (BMAS) is somewhat uniform in universities in Nigeria. The tables below show the sample of the computer science course outline of universities undergraduate programs.

`

YEAR I

**Table I: 100 LEVEL COURSES**

| COURSES | TITLE | UNITS |
|---|---|---|
| CSC 101 | Introduction to Computer Science | 3 |
| CSC 102 | Introduction to Problem Solving | 3 |
| MAT 101 | General Mathematics I | 3 |
| MAT 102 | General Mathematics II | 3 |
| MAT 103 | General Mathematics III | 3 |
| PHY 101 | General Physics I | 3 |
| PHY 102 | General Physics II | 3 |
| PHY 105 | General Physics III | 1 |
| BIO 101 | General Biology I | 3 |
| CHM 101 | General Chemistry I | 3 |
| GES 101 | Use of English | 2 |
| LIB 101 | Library Skills | 1 |

Electives: 6 units to be selected from Mathematics and Physics Courses

YEAR II

**Table II: 200 LEVEL COURSES**

| COURSES | TITLE | UNITS |
|---|---|---|
| CSC 201 | Introduction to Computer Science | 3 |
| CSC 202 | Introduction to Problem Solving | 3 |
| CSC 218 | General Mathematics I | 3 |
| CSC 204 | General Mathematics II | 3 |
| CSC 205 | General Mathematics III | 3 |
| CSC 208 | General Physics I | 3 |
| CSC 212 | General Physics II | 3 |
| MAT 218 | General Physics III | 3 |
| PHS 201 | General Biology I | 3 |
| CSC 299 | General Chemistry I | 3 |
| EPS I | Use of English | 2 |
| GES 201 | Library Skills | 2 |

Electives: 8 units to be selected from MATH 204, Linear Algebra I ( 3 units)
MATH 205, Linear Algebra II (3 units)
PHS 201 Modern Physics ( 3units)
And Statistics courses

YEAR III

**Table III: 300 LEVEL COURSES**

| COURSES | TITLE | UNITS |
|---|---|---|
| CSC 301 | Structured Programming | 3 |
| CSC 302 | Object-Oriented Programming | 3 |
| CSC 310 | Algorithms and Complexity Analysis | 3 |
| CSC 305 | Operating Systems II | 3 |
| CSC 314 | Architecture and Organization I | 3 |
| CSC 315 | Architecture and Organization II | 3 |
| CSC 304 | Data Management I | 3 |
| CSC 316 | Compiler Construction I | 3 |
| CSC 321 | System Analysis and Design | 3 |
| CSC 332 | Survey of Programming Language | 4 |
| CSC 333 | Computational Science & Numerical Methods | 3 |
| CSC 308 | Formal Methods and Software Development | 3 |
| CSC 399 | Industrial Training II | 3 |
| EPS 301 | Entrepreneurship Studies II | 2 |

Electives

| | | | |
|---|---|---|---|
| 6 Units from | CSC 331 - Operation Research | - | 3 |
| | CSC 334 - Numerical Analysis | - | 3 |
| | CSC 335 – Statistical Computing | - | 3 |
| | CSC 306 – Theory of Computing | - | 3 |

YEAR IV

**Table IV: 400 LEVEL COURSES**

| COURSES | TITLE | UNITS |
|---|---|---|
| CSC 403 | Software Engineering | 4 |
| CSC 404 | Data Management II | 3 |
| CSC 421 | Net-Centric Computing | 3 |
| CSC 401 | Organization of Programming Languages | 3 |
| CSC 411 | Artificial Intelligence | 3 |
| CSC 441 | Human Computer Interface | 2 |
| CSC 423 | Computer Networks/Communications | 3 |
| CSC 499 | Project | 6 |

# CHAPTER THREE

# METHODOLOGY

## 3.0 Introduction

This chapter emphasizes on the software development life cycle of this project. The methodology, requirements, system design and development tools will all be would be analyzed and given a diagrammatical expression. The SDLC outlines phases and steps which would be taken in the completion of the project. The system analysis and design is a phase in SDLC. (Hwari and Jain, 2012). During the phase of system analysis, the current system is analyzed closely to find out loopholes in it methods of operation and problems which it poses to clients making use of it. This phase of system analysis in the system development life cycle (SDLC) is usually carried out by a system analyst. The work of the system analyst is to provide the programmers with the specifications and requirement of the system that is about to be build, the programming language to be used and to provide other necessary things to aid the building of the system.

In this chapter, the problem is well understood and possible ways of solving it is brought to light. In the midst of this rundown of programs development phases, the analysis of the system and design phase stands to be the most imperative since it centers around correct documentation of all techniques and analyzes requirements in the format of elaborate diagrams (Ghezi et al, 2002).

System analysis is aimed at understanding the problem which we about to solve in an attempt to solve the problem.

System analysis and design is necessary because it involves

- The specification and requirement of the new system are outlined in details
- Analysis of requirement
- The design of the system
- Diagrams in the form of database design, entity relationship diagram ,data flow diagram (DFD) and Unified Modeling Language.

### 3.1 Software Development Methodology

The waterfall software development model is the software development methodology used for the project.

The Waterfall development model was chosen for this research since all the requirements are known at the start of the system development and the requirements are not subject to change as it is a chronological model in which a development stage must be completed before proceeding to the next stage of development. The waterfall model is the first process model that was introduced. It is a simple model which is also easy to use and understand. The previous phase must be completed before progressing to the next phase. The waterfall software development model is a linear sequential life cycle model that is, it has a linear sequential flow.

### 3.2 System Analysis

This model is a web-based application for receiving feedbacks from the industry to the computer science department. The industry can make suggestions, share their needs and also share information recent findings. The Computer Science department can receive feedback from the industry and reflect in the curriculum. The curriculum is then applied in the department to aid students upon graduation and in the real world. This has not been built before so what have been analyzed is based on the articles that has released on the subject.

## 3.3 Restatement of Project Objectives

The aim of this project is to create a web-based application for computer science department to be able to evaluate their curriculum with the suggestions of the industry.

The objectives are to:

- Design a web-based chat application for the industry and computer science department
- Develop the web-based chat application for the industry and computer science department
- Evaluate the web-based chat application for the industry and computer science department

## 3.4 System Requirements

System requirements are descriptions of functionalities or services offered by the system and its operational limitation the requires of customers for a system that helps provide solution to some problems (Akinde, 2012). There are two main types of requirements that need to be met by the system which are the functional requirements and the non-functional requirement.

### 3.4.1 Functional Requirements

These are statements of services that the system should provide, the behavior of the system when you input certain parameters and reaction if parameters are not imputed. The functional requirement of the system is listed as follows:

- The system shall allow user to make suggestions
- The system shall allow user to review suggestions

### 3.4.2 Non-Functional Requirements

They define the system properties and constraints on the functions. The non-functional requirements of the system are listed as follows:

- Portability: This means that the system is platform independent.
- Interoperability: This is the property of the system that allows the system to be capable of being used or operated reciprocally.

### 3.4.3 User Requirements

These are the needs and desires of the end user of the system. They describe the activities the user is expected to perform. The user requirements of the system are outlined as follows:

- User shall be able to make suggestions
- Administrator shall be able to review suggestions

### 3.4.4 Software Requirements

- Virtual Server: Xampp version 1.7.1, Wamp Server
- JavaScript-enabled web browsers: Internet Explorer, Mozilla Firefox and Google Chrome

## 3.5 Development Tools

These are the tools which will be used in the execution of the system and they will be briefly discussed as follows:

- **XAMPP Server (All Platform, Apache, Mysql, Php, Perl)**

XAMPP is a free and open source application stack which can work on cross platforms. XAMPP helps to create and develop your own web server technology. It provides stability and helps to manage data in database since the web application is meant to be always running.

- **DATABASE MANAGEMENT SYSTEM: MySQL**

MySQL is an open source relational database management system. Its acronym stands for Structured Query Language. MySQL helps to achieve the highest level of scalability, security, reliability and uptime. It helps in better management of data because of the integration and its association of database tables.

### 3.5.1 Programming Languages

The programming languages that would be used in building this system includes PHP, MySQL, JavaScript, CSS, HTML.

- **PHP**

PHP is an acronym for Hypertext Preprocessor. It is one of the major programming tools that would be used in the building of the system. It is a server side scripting language designed for web development but also used as a general purpose programming language.

- **HTML**

HTML is an acronym for Hypertext Markup Language. It is used for creating web pages and web applications. It is used in Cascading Style Sheets (CSS) and JavaScript (JS). HTML is sometimes referred to as the foundation of all web pages.

- **JavaScript**

JavaScript is a programming language that authorizes you to execute complicated things on web pages. It was developed as a scripting language that web server administrators could use to manage the server and connect its pages to other services.

- **CSS**

Cascading Style Sheet (CSS) is used to ascertain the form, blueprint, demonstration and appearance of web page components.

### 3.5.2 CASE TOOLS

Computer Aided Software Engineering tools is the use of software tools to assist in the development and maintenance of software. Some CASE tools that would be used in the building of this system include Enterprise Architect, MySQL Workbench, Edraw Max and Microsoft Project.

- **Enterprise Architect**

Enterprise Architecture enables you to conduct enterprise analysis, design, planning and implementation, debug, run and execute utilize scripts all from among the enterprise architect development environment.

- **Microsoft Project**

Microsoft Project that aids the management of a project. It helps a project manager to plan, assign resources to tasks and to determine the approximate end of a project. It can also be used to manage the budget of the project and keep track of progress.

## 3.6 SYSTEM DESIGN

The phase involves the use of tools such as the Data Flow Diagram (DFDs), Data Dictionary, Entity Relationship Diagram (ERDs), Unified Modeling Language (UML) and so on.

### 3.6.1 Unified Modeling Language (UML)

The Unified Modeling Language (UML) is a graphical language for visualizing, specifying, constructing and documenting the artifacts of a software-intensive system as stated by The Object Management Group (OMG). Unified modeling language is a demonstrative tool exemplifying extensively the structure of a software-intensive system, as well as users and behavioral interactions with the software system. Some example of UML diagrams include Use Case, Class

Diagram, Component Diagram and others. The UML diagram that is used in this project is the USE case diagram.

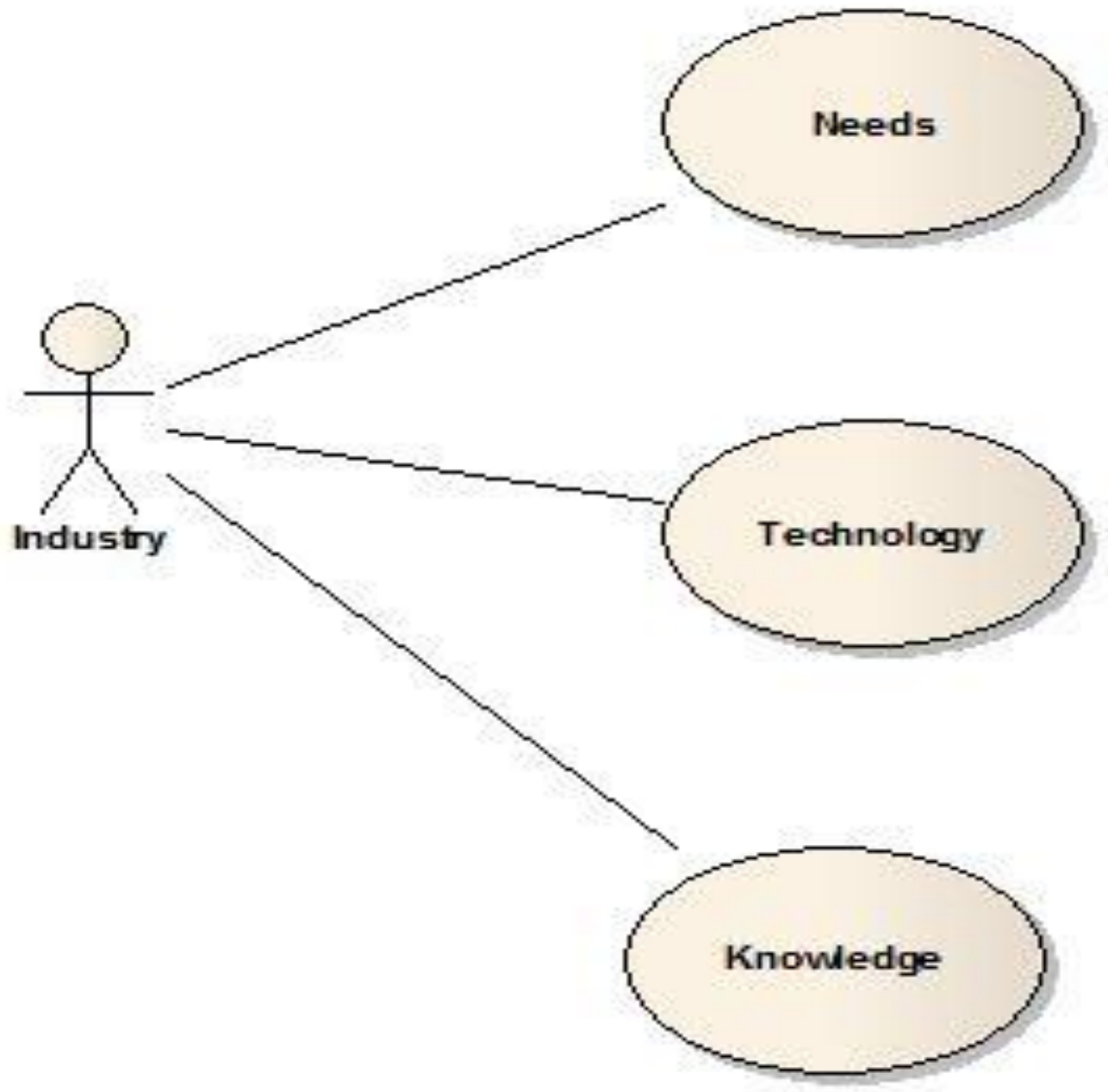

Figure 3.1 – Unified Modeling Language use case diagram

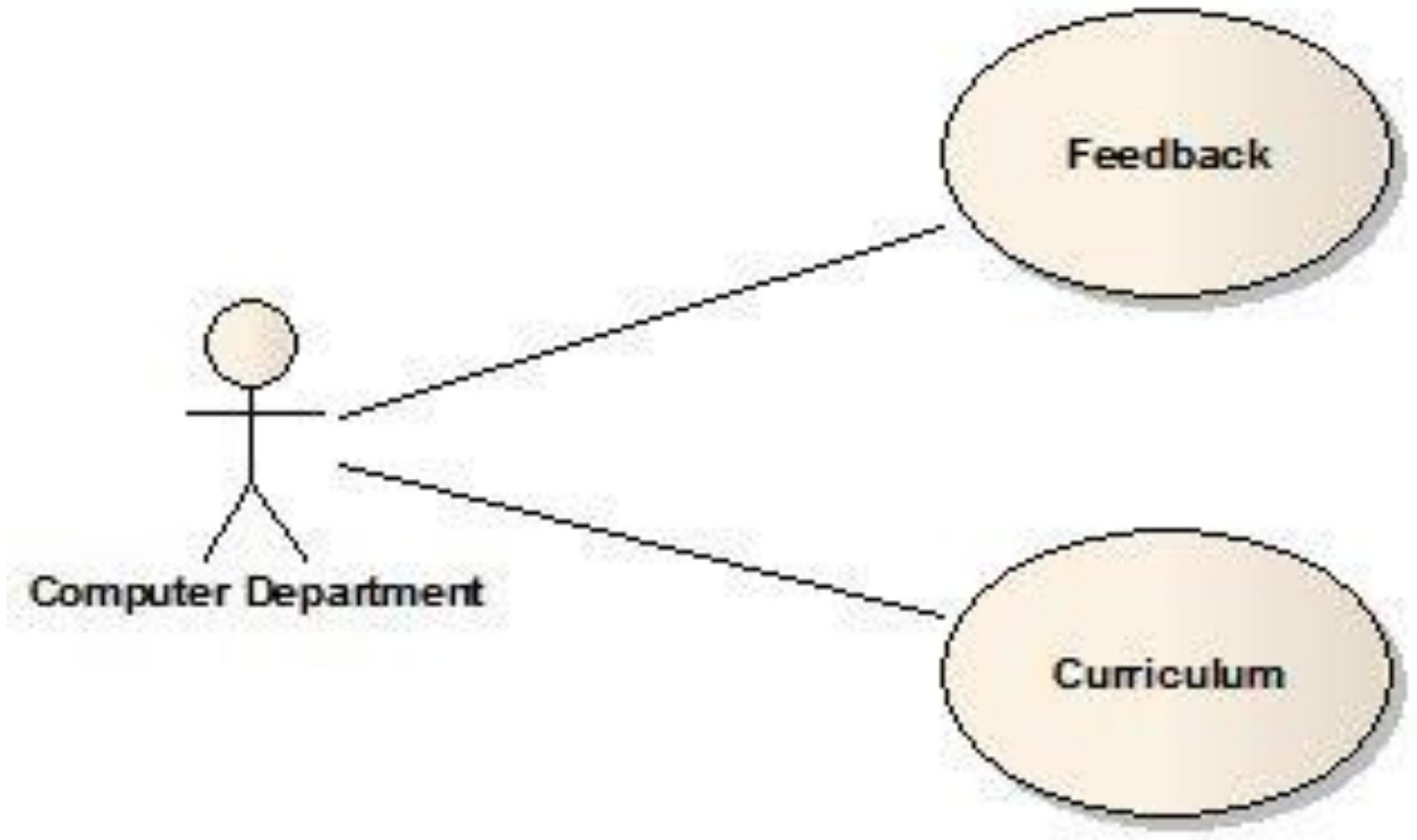

Figure 3.2 – Unified Modeling Language use case diagram

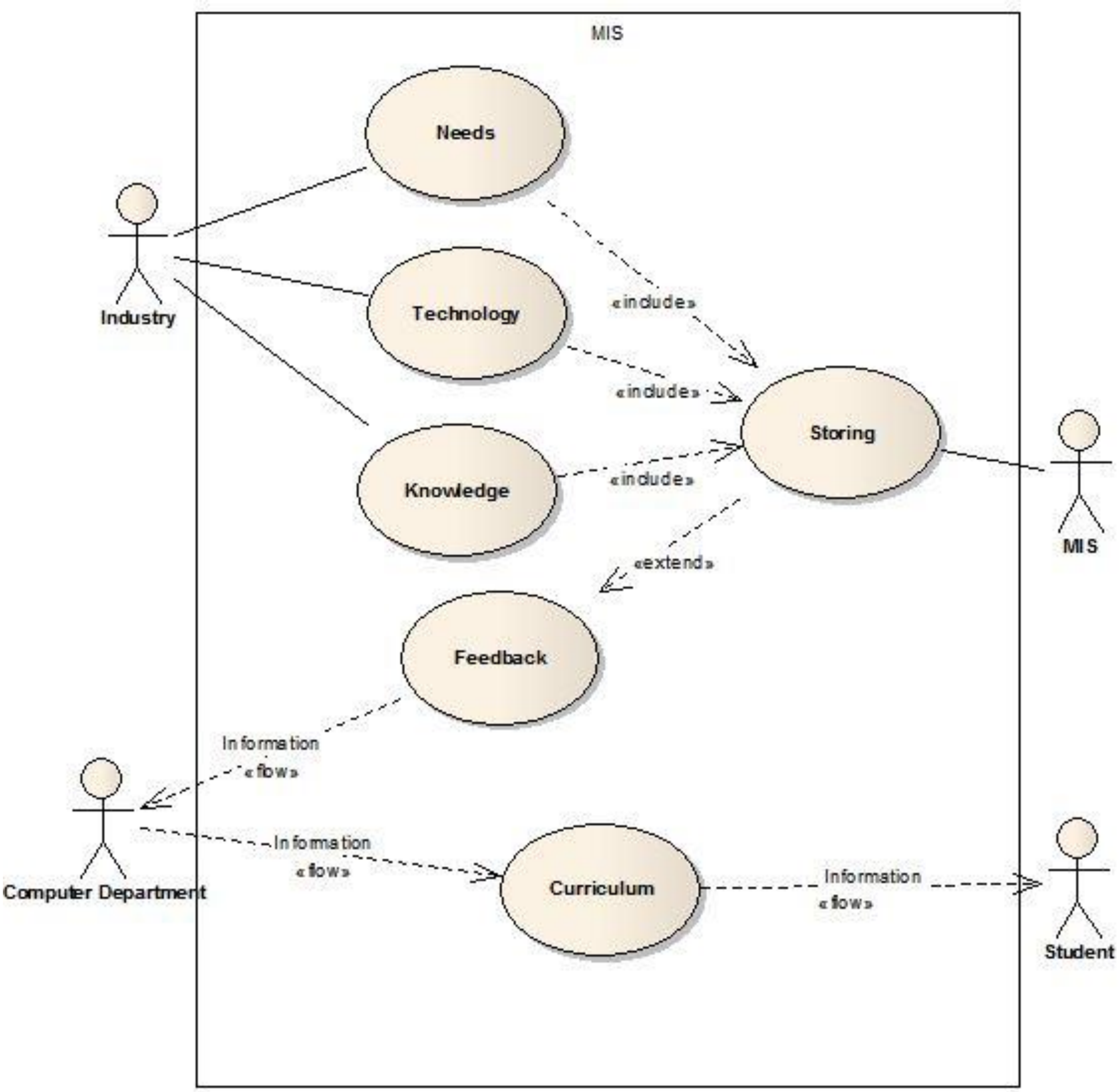

Figure 3.3 – Unified Modeling Language use case diagram

### 3.6.2 Data Flow Diagram

The data flow diagram outlines the information flow in a system. It uses shapes and symbols to outline, in terms of inputs and `outputs, it is how data is processed. It shows the source of information and the destination of the information. It can also be said to be the flow of data through an information system using graphical representation.

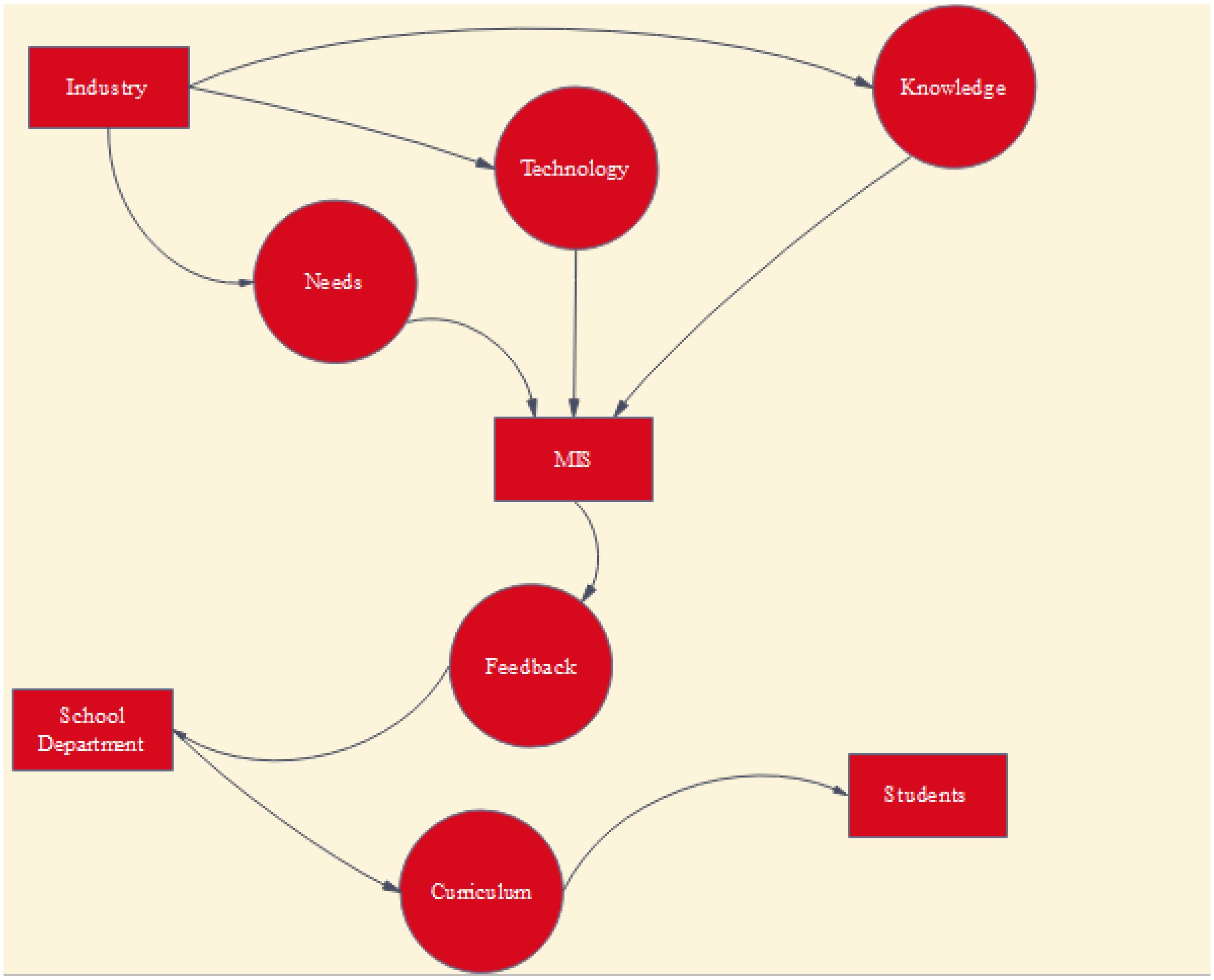

Figure 3.4 – Data Flow Diagram

### 3.6.3 Entity Relationship Diagram

It is a graphical representation that used for the design and analysis phase in the Software Development Life Cycle (SDLC) to describe entities of a software-intensive system from a top-down perspective. It describes how entities relates with one another or the relationship between entities, objects or events within that system.

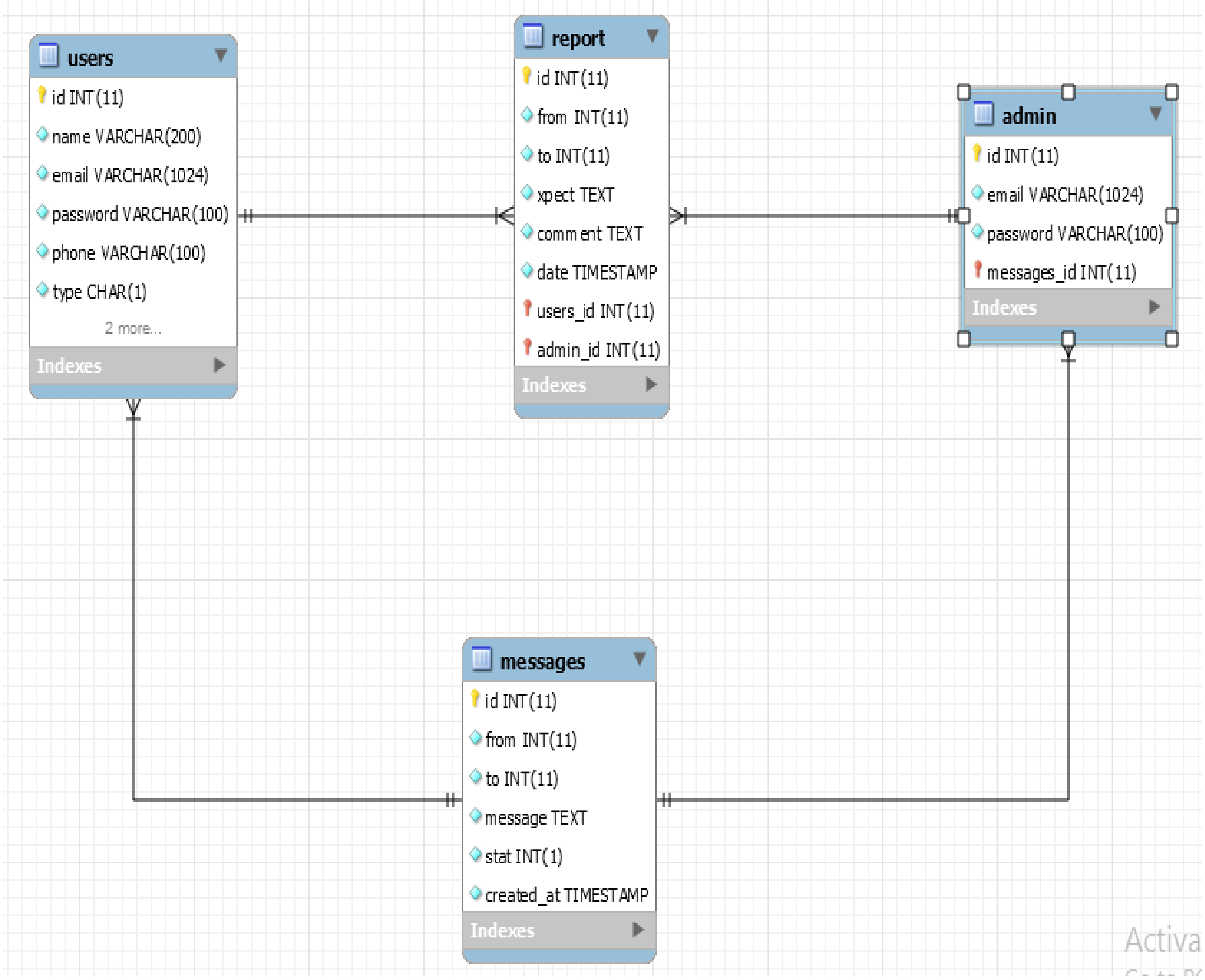

# CHAPTER FOUR

# SYSTEM IMPLEMENTATION & TESTING

## 4.1. INTRODUCTION

This chapter deals on the implementation of the system. The software is also tested in this chapter, to check for errors and how effective the system is interacting with its other components.

## 4.2. SYSTEM COMPONENTS

The developed system leverages on web development tools and techniques in implementing a state-of-the-art, use friendly and cost effective system. The system was designed for the primary purpose of facilitating easy exchange collaboration between companies and university.

The testing of the components was based on the database and the user interface of the system.

The system must be tested to check for errors in other for corrections to be made.

### 4.2.1 USER INTERFACE

This sector discusses the various features of the developed system after integration testing, before getting to this phase, each of the component of the system have been tested and found to work effectively. In subsequent sub-divisions, we will be explaining how the various modules works

The various modules that makeup the propose system are discussed below:

1. Register page: Both the companies and schools that want to use the system must register.

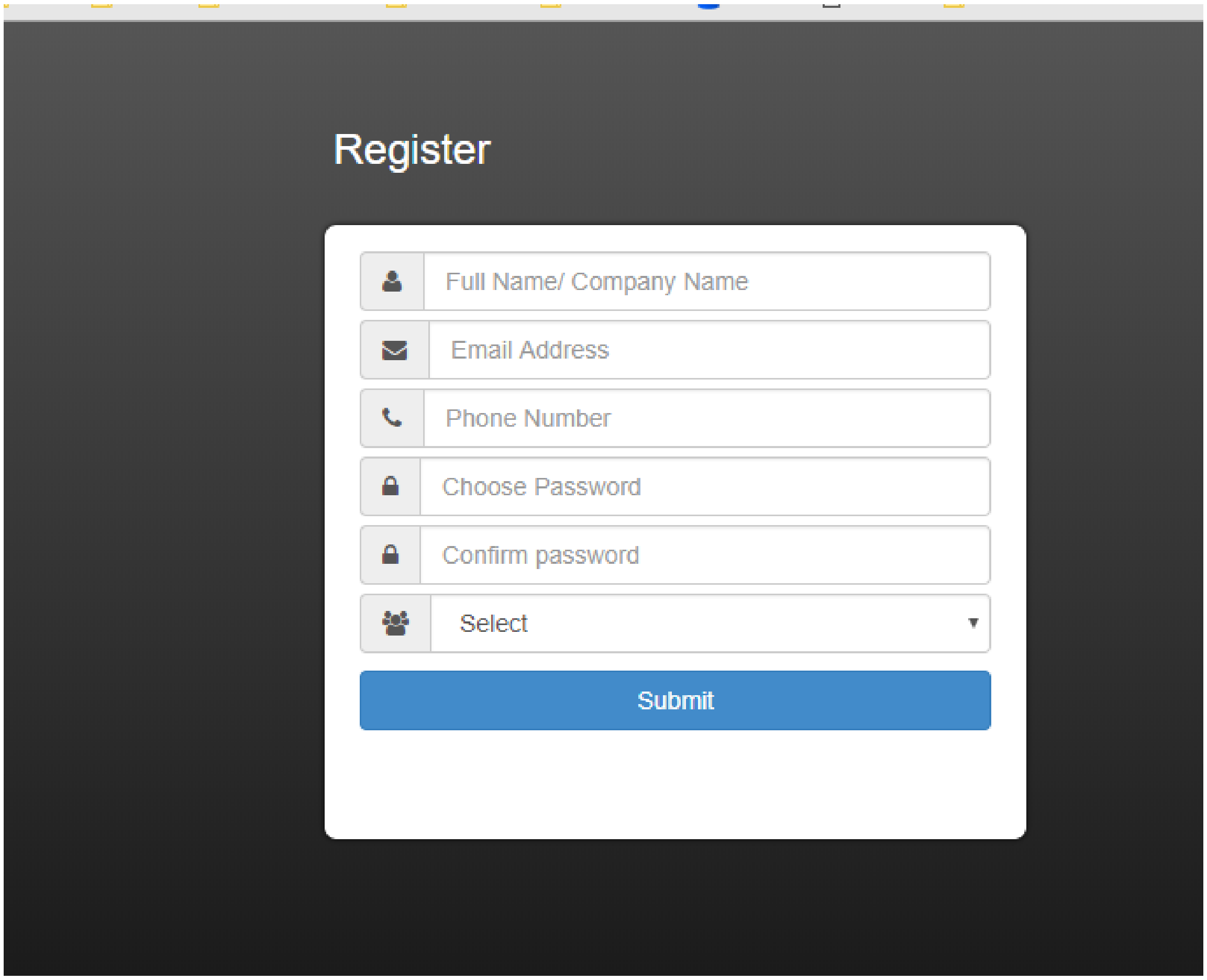

**Figure 4.1 Register page**

2. **Login Page :** Registered company or school, and admin must sign in with valid registered details before they can gain access to the system

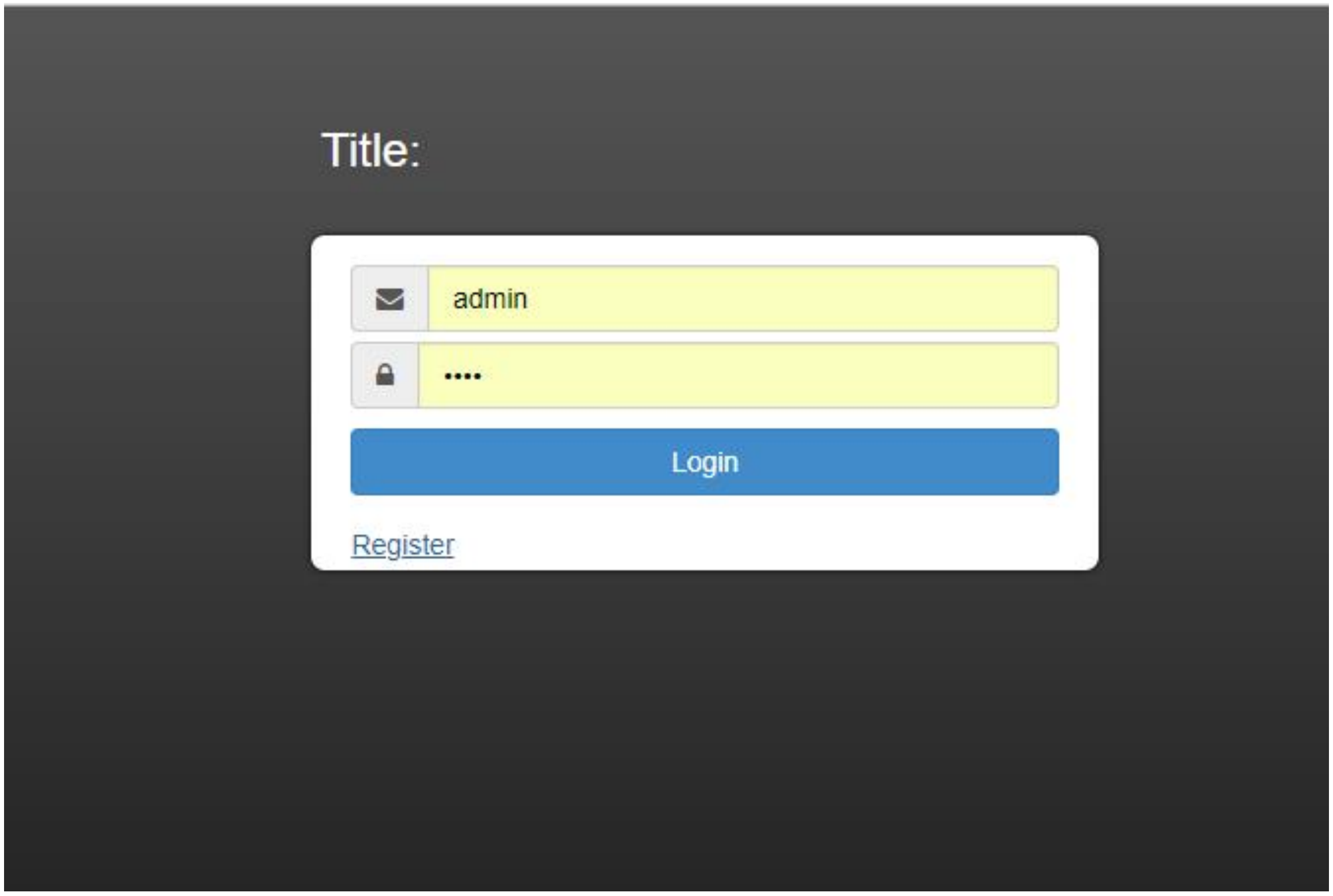

**Figure 4.2 Login Page**

3. Message List: Both Company and School can view the list of messages sent to them from this page.

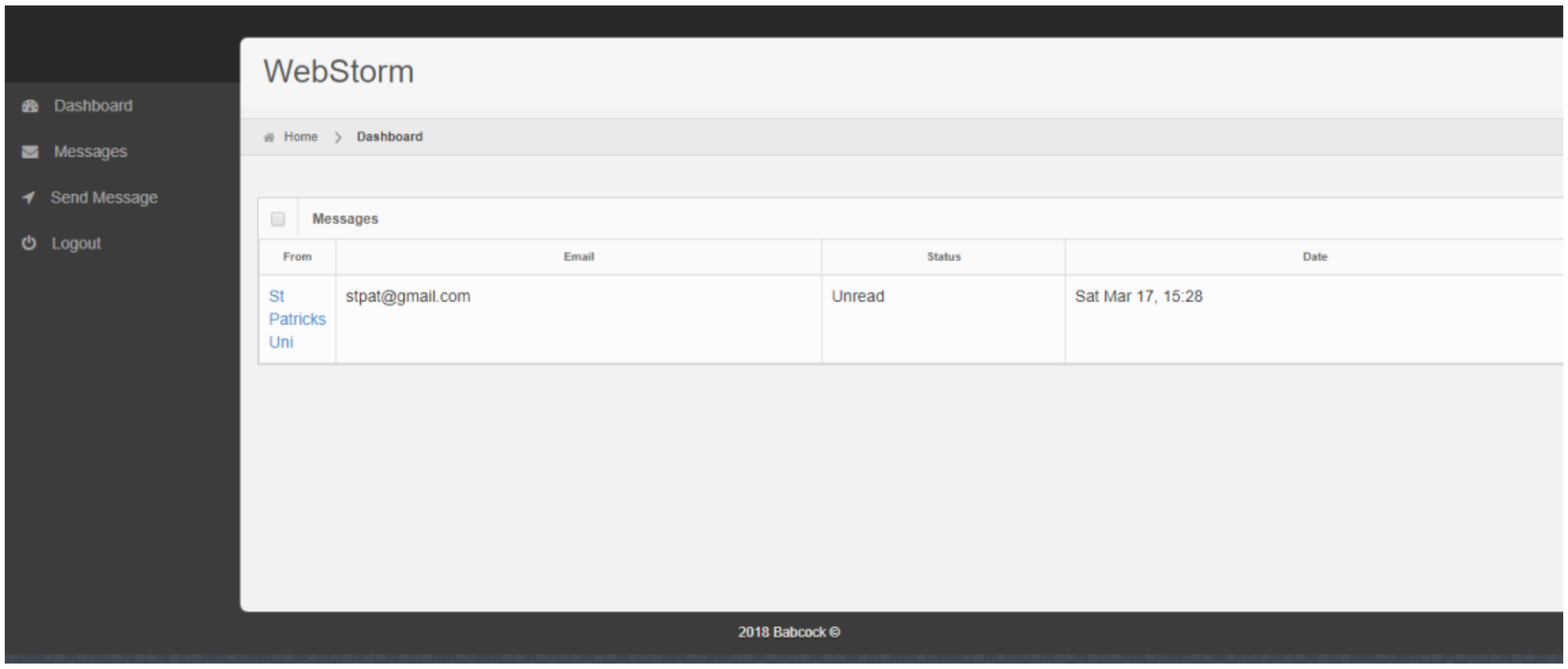

**Figure 4.3 Message List Page**

4. **Send Message Page:** Company can send message to school from this page and vice versa.

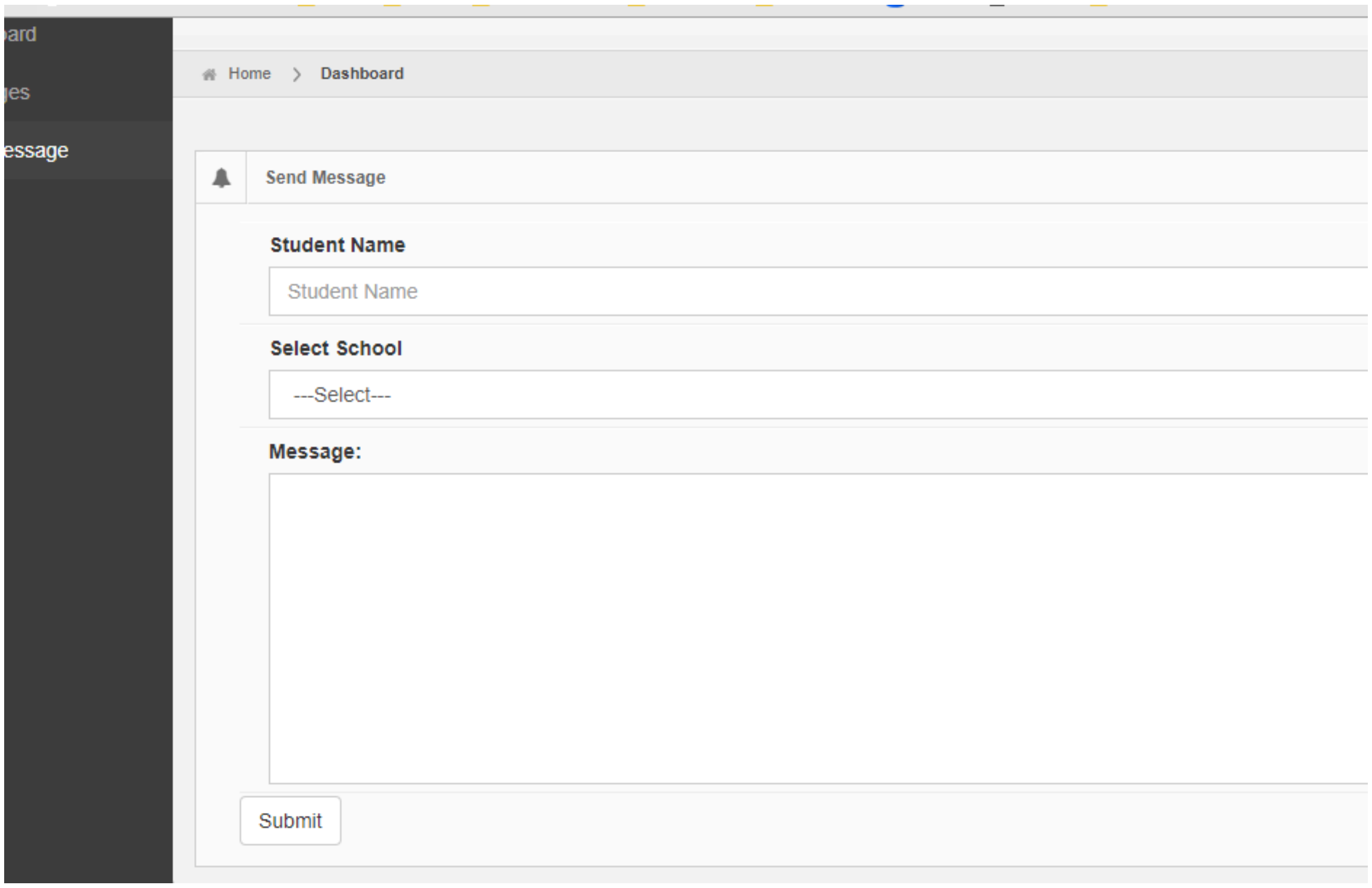

**Figure 4.4 Message Page**

5. **Send Report Page:** Company can send student report to school from this page by entering all the required details and selecting the name of the school.

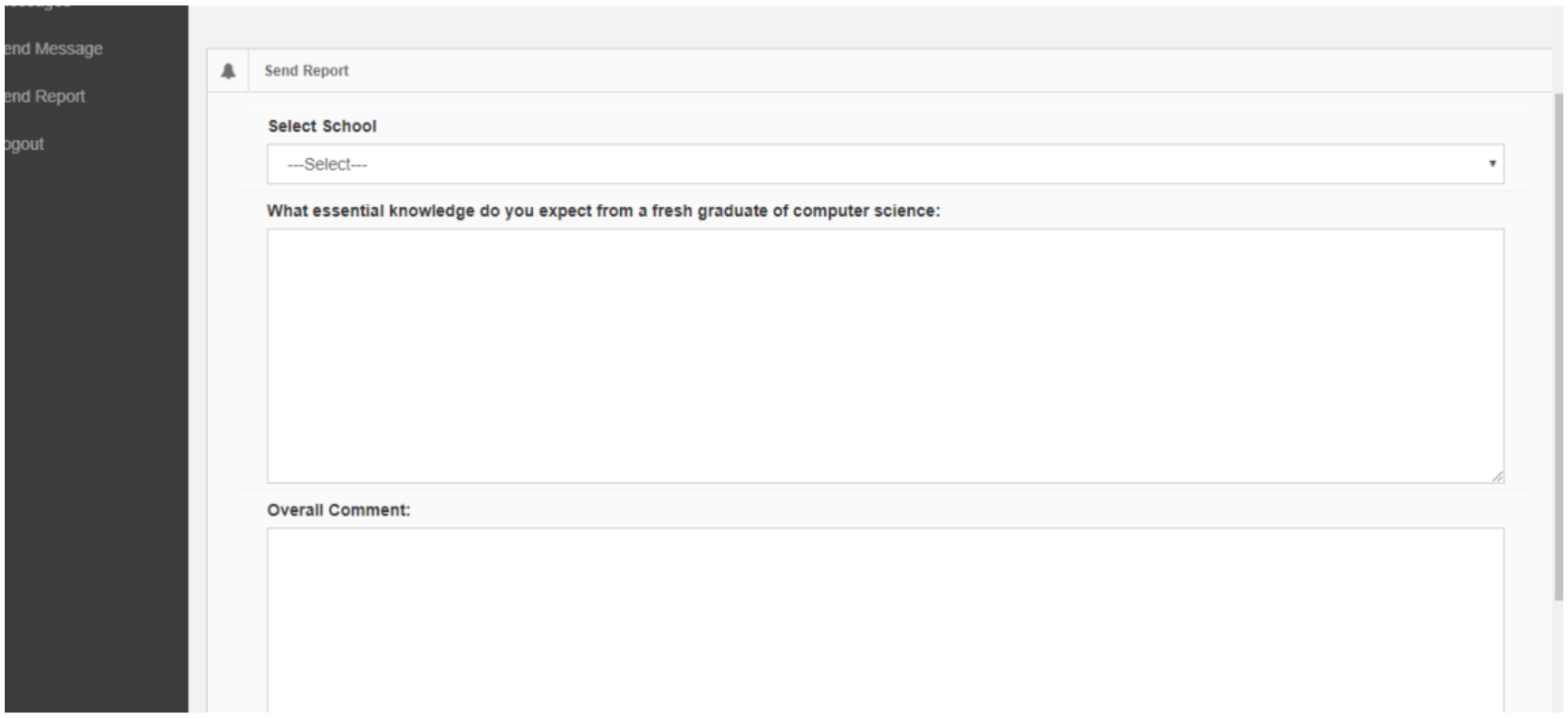

**Figure 4.5 Send Report Page**

6. **View Report:** Schools can view the report sent to them by company on this page.

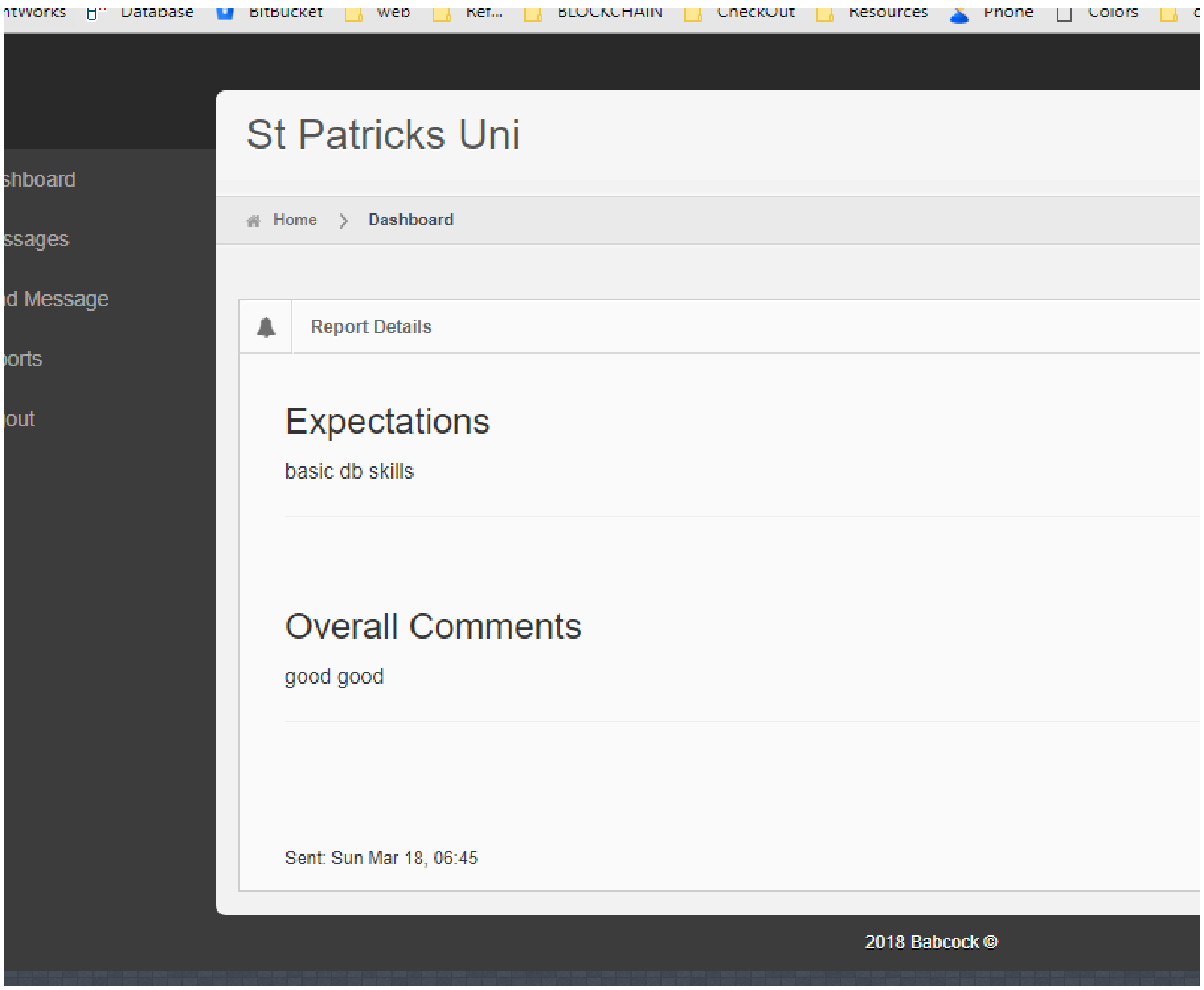

**Figure 4.6 View Report Page**

### 4.2.2 DATABASE:

The most crucial part of a database is the table because it is the actual content-holder created to retain a set of related data. Before any data is stored, the Database Management System (DBMS) confirms whether all rules governing the table have been implemented. Each of the tables in the database was tested and found to meet all expectations. Screenshots of the various tables are shown below.

### DATABASE COMPONENTS

All the tables that made up the system database are listed below:

1.Admin Table: Stores admin details.

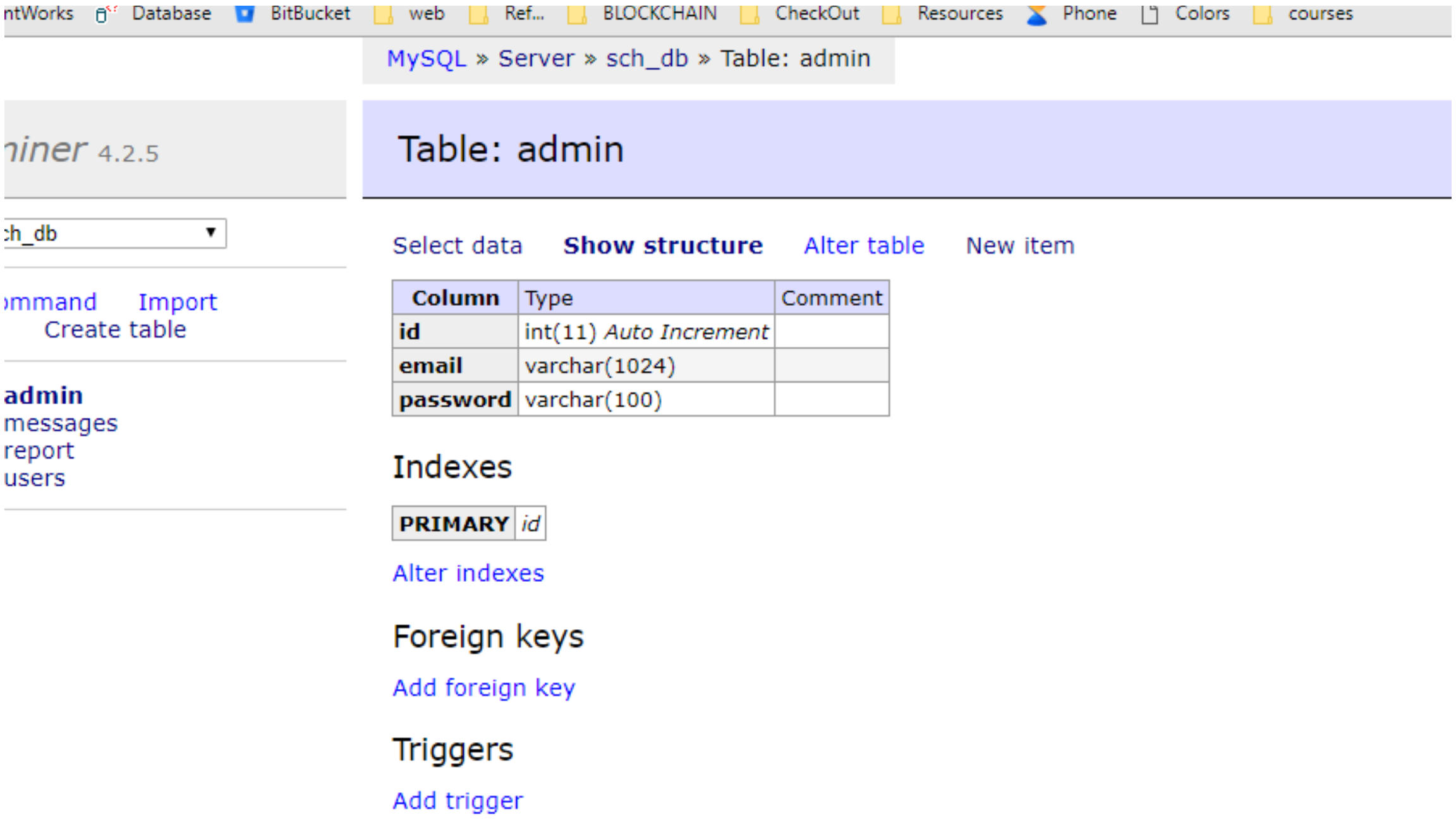

Figure 4.7 Admin table

2. Message Table ; Keep record of messages between companies and schools

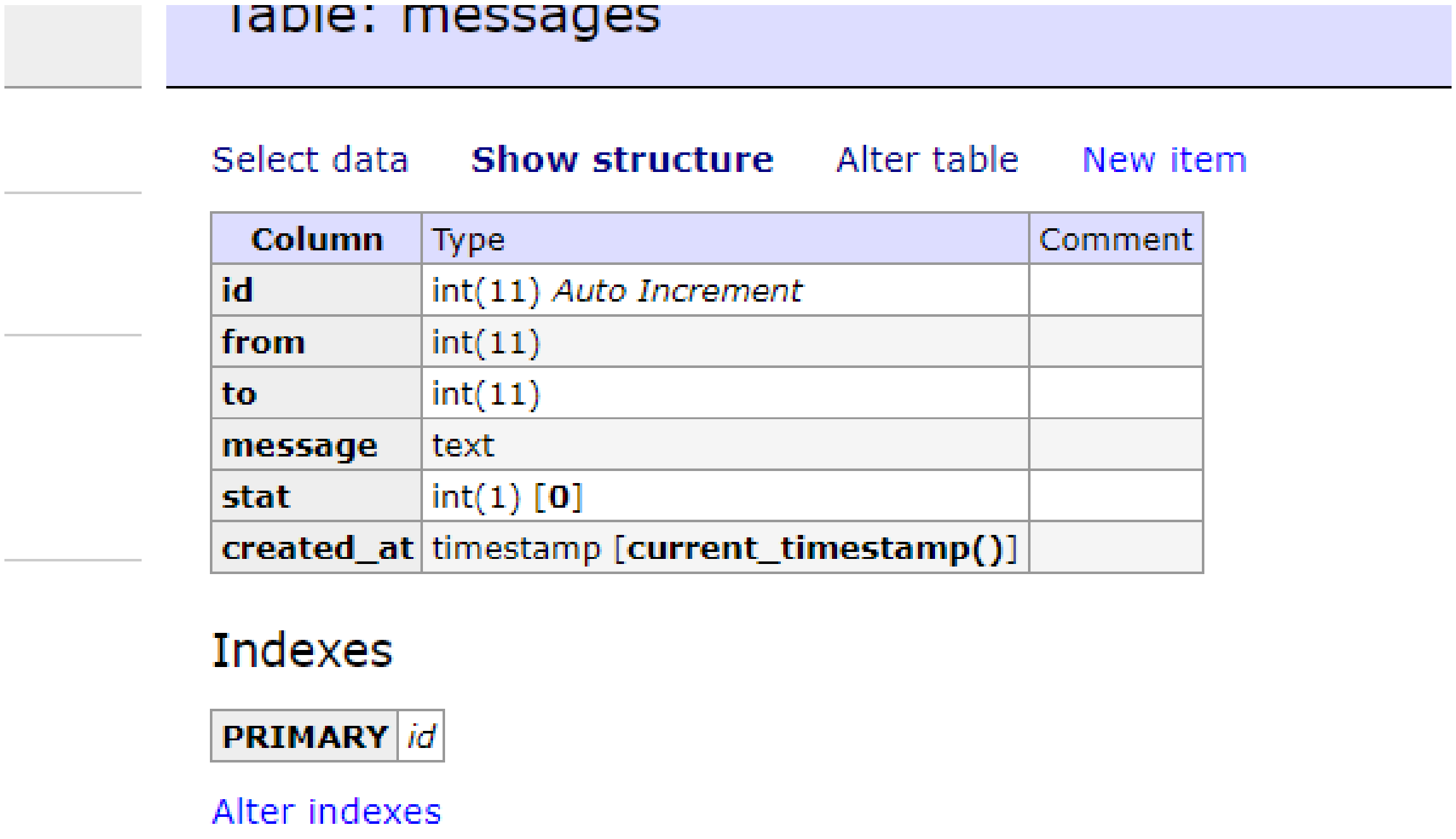

Table: messages

Select data Show structure Alter table New item

| Column | Type | Comment |
|---|---|---|
| **id** | int(11) *Auto Increment* | |
| **from** | int(11) | |
| **to** | int(11) | |
| **message** | text | |
| **stat** | int(1) [**0**] | |
| **created_at** | timestamp [**current_timestamp()**] | |

Indexes

| | |
|---|---|
| **PRIMARY** | *id* |

Alter indexes

Figure 4.8 Message table

3 User Table: Stores both public and authentication details of both companies and schools.

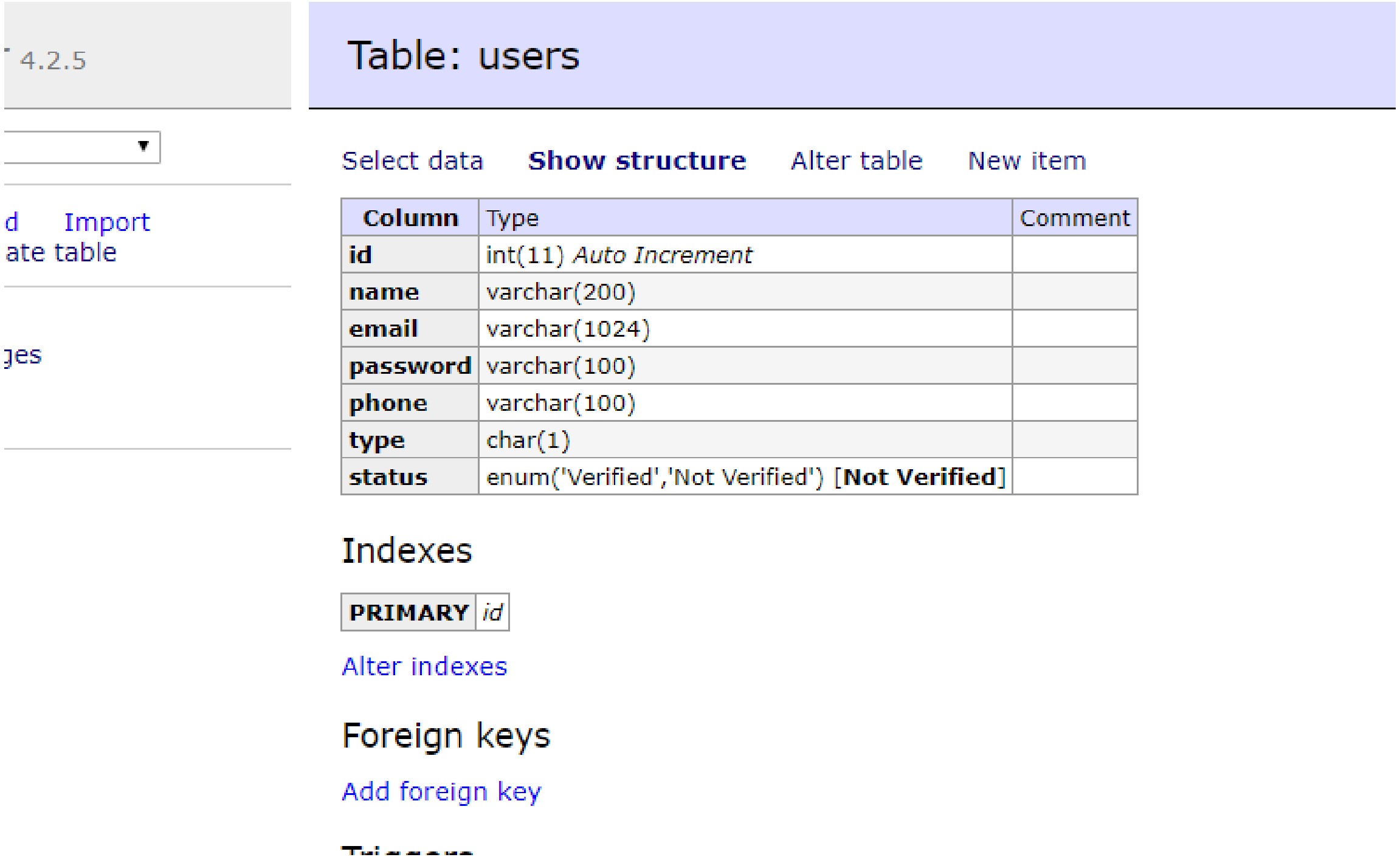

Table: users

Select data Show structure Alter table New item

| Column | Type | Comment |
|---|---|---|
| **id** | int(11) *Auto Increment* | |
| **name** | varchar(200) | |
| **email** | varchar(1024) | |
| **password** | varchar(100) | |
| **phone** | varchar(100) | |
| **type** | char(1) | |
| **status** | enum('Verified','Not Verified') [**Not Verified**] | |

Indexes

| | |
|---|---|
| **PRIMARY** | *id* |

Alter indexes

Foreign keys

Add foreign key

Figure 4.9 Users table

4 Report Table: Keep records of students report send to school by company

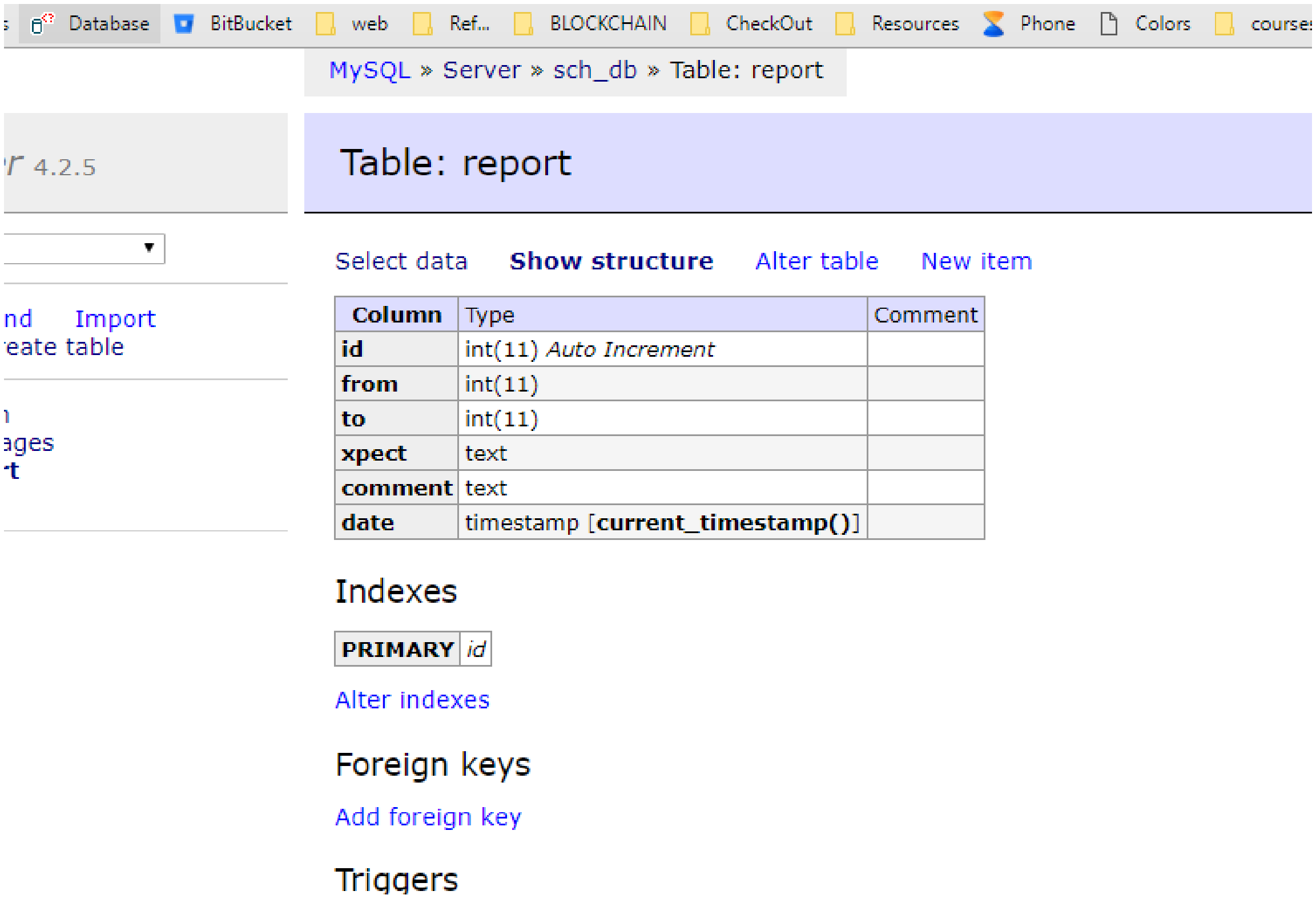

| Column | Type | Comment |
|---|---|---|
| **id** | int(11) *Auto Increment* | |
| **from** | int(11) | |
| **to** | int(11) | |
| **xpect** | text | |
| **comment** | text | |
| **date** | timestamp [**current_timestamp()**] | |

Figure 4.91 Report table

## 4.3 SYSTEM REQUIREMENT

Based on the testing that has been carried out, the minimum software and hardware requirements carried out are given as fellows;

### 4.3.1 Hardware Requirements

The minimum hardware requirements include:

1. Hard disk capacity: 5 GB

2. Main processor: Pentium IV

3. Random access memory: 512 MB

### 4.3.2 Software Requirements

The minimum software requirements for the system to be accessible include:

1. JavaScript-enabled web browsers: Mozilla Firefox (most suitable), Internet Explorer, Google Chrome, and Opera-mini

2. Virtual server: Xampp version 1.7.1

3. Operating system: Windows (98, 2000, ME, NT, XP, Vista, 7), Linux, Mac OS

## 4.4 Risk Identification, Category, Component and Mitigation

A risk is the possibility that an undesirable event could happen. Software risks may be characterized as follows:

i. Uncertainty - the risk may or may not happen i.e. there are no 100% probable risks.

ii. Loss - if the risk becomes a reality, unwanted consequences or losses will occur.

### 4.4.1 Risk Components

There are four different risk components:

**i.** **Performance Risk**: It is the degree of uncertainty that the product will meet its requirements and be fit for its intended use.

**ii.** **Cost Risk:** It is the degree of uncertainty that the project budget will be maintained.

**iii.** **Support Risk**: The degree of uncertainty that the resultant software will be easy to correct, adapt, support, and enhanced

**iv.** **Schedule Risk**: The degree of uncertainty that the project schedule will be maintained and that the product will be delivered on time.

### 4.4.2 Impact

This refers to what damage the risk identified can do to the project. It is classified into Negligible, Marginal, Critical and Catastrophic.

### 4.4.3 RISK CATEGORIES

Risks fall under the following categories:

i. **Project Risk**: These types of risk threaten project plan. They help identify potential problems with schedule, budget, resource allocation and requirement problems.

ii. **Technical Risk**: These threaten the quality and timeliness of the software to be produced.

iii. **Business Risk:** These threaten the viability of the software to be built.

### 4.4.4 Risk Mitigation, Monitoring & Management (RMMM)

This refers to the solution put to correct the risk at hand. A good RMMM should try to avoid the risk, monitor and also manage it.

Table 1: Risk Mitigation, Monitoring & Management.

| Risk | Risk Component | Impact | Risk Category | Probability | RMMM |
|---|---|---|---|---|---|
| The technology to be used for the project is new to most members of the group. | Schedule risk | Marginal | Project risk<br><br>Technical risk | 30% | Project members were put under tight schedule to learn and implement the technology after class hours. |
| Deadline for project submission will be tightened. | Performance risk<br>Schedule risk | Critical | Project risk<br><br>Technical risk | 40% | More attention would be put towards building the application. |
| | | | Business risk | | |
| Requirements might change during the building process. | Performance risk Schedule risk | Marginal | Project risk<br><br>Technical risk | 30% | More time would invested in gathering all the possible requirements and perfect understanding of the project before development. |

# CHAPTER FIVE

## SUMMARY, CONCLUSION AND RECOMMENDATION

## 5.1 Summary

The new world technology today is changing rapidly and the use of the industry curriculum system is a way of making sure it is known throughout the world. The aim and objectives of this project was to design a web based chat application for the industry and computer science department, develop a web based chat application for the industry and computer science department and evaluate the web based chat application for the industry and computer science department which was achieved in this project.

The chapter one of this project basically includes the introduction of the project, the aim and objectives, the scope of study and the problem is stated which the solution has been provided. The chapter two gives a brief background of previous related articles. The chapter three discusses the methodology used in the project. It discusses the development tools, software development life cycle and the design of the software. The chapter four discusses the implementation of the things discussed in chapter three, It also discusses the working and screenshots of the software. It talks about component testing, database testing involving data validation and integrity, system documentation, interface testing.

From the problem statement the major problem that is been faced is the lack of communication between the computer science department and the industry. The software developed will helped to bridge that gap in educational institutions and will also be advantageous to the industry because they will have more knowledgeable graduates for employment.

## 5.2 Conclusion

Due to the research carried out, the developed software is assumed to be unique and is assumed to be the first of its kind. This research presents the computer science department with a way of improving their curriculum by aligning it with the needs of the industry. Based on the previous articles that were reviewed, the project was developed to suit the suggestion and recommendations that were made.

## 5.3 Recommendation

The system is not a perfect system, so there is room for improvement to provide the greatest value of the system to the industry and the computer science department. Some of these recommendation include:

1. Create awareness of the software to companies and universities
2. Implement the suggestions made by the industry in the computer science curriculum
3. Use of this software in universities across Nigeria
4. Use of this not just in the computer science field but in other field of studies

**SOURCE CODE**

```
//index
<!DOCTYPE html>
<html lang="en">
<head>
        <title>Dashboard | Login</title>
                <meta charset="UTF-8" />
        <meta name="viewport" content="width=device-width, initial-scale=1.0" />
                <link rel="stylesheet" href="css/bootstrap.min.css" />
        <link rel="stylesheet" href="css/font-awesome.css" />
        <link rel="stylesheet" href="css/unicorn-login.css" />
                <script type="text/javascript" src="js/respond.min.js"></script>

        </head>    <body>
        <div id="container">
            <div id="logo">
                <h3 style="color: white">Title:</h3>
            </div>
            <div id="here"></div>
            <div id="loginbox" style="height: 170px">
                <form action="core/login.php" method="post">
                                <p>  </p>
                    <div class="input-group input-sm">
```

```
          <span class="input-group-addon"><i class="fa fa-envelope"></i></span><input
class="form-control" type="text" name="email" id="email" placeholder="Email Address" />
        </div>

        <div class="input-group">
          <span class="input-group-addon"><i class="fa fa-lock"></i></span><input
class="form-control" type="password" name="password" id="password"
placeholder="Password" />
        </div>
        <div id="profi" class="form-actions clearfix">
          <button id="subo" type="submit" class="btn  btn-block btn-primary btn-default"
value="">Login</button>
        </div>
        <div class="input-group">
          <a href="register.php"> Register</a>
        </div>
      </form>
    </div>
  </div>

  <script src="js/jquery.min.js"></script>
  <script src="js/jquery-ui.custom.min.js"></script>
  <!-- <script src="js/unicorn.login.js"></script>  -->
```

```
        <script src="js/custom.js" type="text/javascript"></script>
    </body>
</html>
//register.php
<?php include "core/controller.php";?>
<!DOCTYPE html>
<html lang="en">
<head>
        <title>Dashboard | Register</title>
                        <meta charset="UTF-8" />
        <meta name="viewport" content="width=device-width, initial-scale=1.0" />
                        <link rel="stylesheet" href="css/bootstrap.min.css" />
        <link rel="stylesheet" href="css/font-awesome.css" />
        <link rel="stylesheet" href="css/unicorn-login.css" />
                        <script type="text/javascript" src="js/respond.min.js"></script>

        </head>    <body>
        <div id="container">
            <div id="logo">
                <h3 style="color: white">Register</h3>
            </div>
            <div id="here"></div>
            <div id="loginbox">
```

```
        <form action="core/register.php" method="post" >
                        <p> </p>
          <div class="input-group input-sm">
            <span class="input-group-addon"><i class="fa fa-user"></i></span><input
class="form-control" type="text" name="fname" id="fname" placeholder="Full Name/
Company Name" />
          </div>
          <div class="input-group input-sm">
            <span class="input-group-addon"><i class="fa fa-envelope"></i></span><input
class="form-control" type="text" name="email" id="email" placeholder="Email Address" />
          </div>
          <div class="input-group input-sm">
            <span class="input-group-addon"><i class="fa fa-phone"></i></span><input
class="form-control" type="number" name="phone"  placeholder="Phone Number" />
          </div>
          <div class="input-group">
            <span class="input-group-addon"><i class="fa fa-lock"></i></span><input
class="form-control" type="password" name="password" id="password" placeholder="Choose
Password" />
          </div>
          <div class="input-group">
```

```
          <span class="input-group-addon"><i class="fa fa-lock"></i></span><input
class="form-control" type="password" name="password2" id="password2"
placeholder="Confirm password" />
        </div>
        <div class="input-group">
          <span class="input-group-addon">
            <i class="fa fa-users"></i>
          </span>
            <select name="type" id="" required class="form-control">
              <option value="">Select</option>
              <option value="S">School</option>
              <option value="C">Company</option>
            </select>

        </div>
        <div id="profi" class="form-actions clearfix">
          <button  type="submit" class="btn  btn-block btn-primary btn-default"
value="">Submit</button>
        </div>
      </form>
    </div>
  </div>
```

```
        <script src="js/jquery.min.js"></script>

        <script src="js/jquery-ui.custom.min.js"></script>

        <!-- <script src="js/custom.js" type="text/javascript"></script> -->

    </body>

</html>


//login

<?php

if (session_id() == "") session_start(); // Initialize Session data

ob_start(); // Turn on output buffering

?>

<?php include_once "ewcfg10.php" ?>

<?php include_once "ewmysql10.php" ?>

<?php include_once "phpfn10.php" ?>

<?php include_once "admininfo.php" ?>

<?php include_once "userfn10.php" ?>

<?php


//

// Page class

//


$login = NULL; // Initialize page object first
```

```
class clogin extends cadmin {

	// Page ID
	var $PageID = 'login';

	// Project ID
	var $ProjectID = "{A9DCF167-C4A5-4E05-A3AE-CC35287D14E1}";

	// Page object name
	var $PageObjName = 'login';

	// Page name
	function PageName() {
		return ew_CurrentPage();
	}

	// Page URL
	function PageUrl() {
		$PageUrl = ew_CurrentPage() . "?";
		return $PageUrl;
	}
```

```
// Message
function getMessage() {
	return @$_SESSION[EW_SESSION_MESSAGE];
}

function setMessage($v) {
	ew_AddMessage($_SESSION[EW_SESSION_MESSAGE], $v);
}

function getFailureMessage() {
	return @$_SESSION[EW_SESSION_FAILURE_MESSAGE];
}

function setFailureMessage($v) {
	ew_AddMessage($_SESSION[EW_SESSION_FAILURE_MESSAGE], $v);
}

function getSuccessMessage() {
	return @$_SESSION[EW_SESSION_SUCCESS_MESSAGE];
}

function setSuccessMessage($v) {
	ew_AddMessage($_SESSION[EW_SESSION_SUCCESS_MESSAGE], $v);
```

```
	}

	function getWarningMessage() {
		return @$_SESSION[EW_SESSION_WARNING_MESSAGE];
	}

	function setWarningMessage($v) {
		ew_AddMessage($_SESSION[EW_SESSION_WARNING_MESSAGE], $v);
	}

	// Show message
	function ShowMessage() {
		$hidden = FALSE;
		$html = "";

		// Message
		$sMessage = $this->getMessage();
		$this->Message_Showing($sMessage, "");
		if ($sMessage <> "") { // Message in Session, display
			if (!$hidden)
				$sMessage = "<button type=\"button\" class=\"close\" data-
dismiss=\"alert\">&times;</button>" . $sMessage;
```

```
			$html .= "<div class=\"alert alert-success ewSuccess\">" . $sMessage .
"</div>";
			$_SESSION[EW_SESSION_MESSAGE] = ""; // Clear message in
Session
		}

		// Warning message
		$sWarningMessage = $this->getWarningMessage();
		$this->Message_Showing($sWarningMessage, "warning");
		if ($sWarningMessage <> "") { // Message in Session, display
			if (!$hidden)
				$sWarningMessage = "<button type=\"button\" class=\"close\"
data-dismiss=\"alert\">&times;</button>" . $sWarningMessage;
			$html .= "<div class=\"alert alert-warning ewWarning\">" .
$sWarningMessage . "</div>";
			$_SESSION[EW_SESSION_WARNING_MESSAGE] = ""; // Clear
message in Session
		}

		// Success message
		$sSuccessMessage = $this->getSuccessMessage();
		$this->Message_Showing($sSuccessMessage, "success");
		if ($sSuccessMessage <> "") { // Message in Session, display
```

```
				if (!$hidden)
					$sSuccessMessage = "<button type=\"button\" class=\"close\"
data-dismiss=\"alert\">&times;</button>" . $sSuccessMessage;
				$html .= "<div class=\"alert alert-success ewSuccess\">" .
$sSuccessMessage . "</div>";
				$_SESSION[EW_SESSION_SUCCESS_MESSAGE] = ""; // Clear
message in Session
			}

			// Failure message
			$sErrorMessage = $this->getFailureMessage();
			$this->Message_Showing($sErrorMessage, "failure");
			if ($sErrorMessage <> "") { // Message in Session, display
				if (!$hidden)
					$sErrorMessage = "<button type=\"button\" class=\"close\" data-
dismiss=\"alert\">&times;</button>" . $sErrorMessage;
				$html .= "<div class=\"alert alert-error ewError\">" . $sErrorMessage .
"</div>";
				$_SESSION[EW_SESSION_FAILURE_MESSAGE] = ""; // Clear
message in Session
			}
			echo "<table class=\"ewStdTable\"><tr><td><div class=\"ewMessageDialog\"" .
(($hidden) ? " style=\"display: none;\"" : "") . ">" . $html . "</div></td></tr></table>";
```

```
}
var $PageHeader;
var $PageFooter;

// Show Page Header
function ShowPageHeader() {
	$sHeader = $this->PageHeader;
	$this->Page_DataRendering($sHeader);
	if ($sHeader <> "") { // Header exists, display
		echo "<p>" . $sHeader . "</p>";
	}
}

// Show Page Footer
function ShowPageFooter() {
	$sFooter = $this->PageFooter;
	$this->Page_DataRendered($sFooter);
	if ($sFooter <> "") { // Footer exists, display
		echo "<p>" . $sFooter . "</p>";
	}
}

// Validate page request
```

```
function IsPageRequest() {
	return TRUE;
}

//
// Page class constructor
//
function __construct() {
	global $conn, $Language;
	$GLOBALS["Page"] = &$this;

	// Language object
	if (!isset($Language)) $Language = new cLanguage();

	// Parent constuctor
	parent::__construct();

	// Table object (admin)
	if (!isset($GLOBALS["admin"])) {
		$GLOBALS["admin"] = &$this;
		$GLOBALS["Table"] = &$GLOBALS["admin"];
	}
	if (!isset($GLOBALS["admin"])) $GLOBALS["admin"] = &$this;
```

```
		// Page ID
		if (!defined("EW_PAGE_ID"))
			define("EW_PAGE_ID", 'login', TRUE);

		// Start timer
		if (!isset($GLOBALS["gTimer"])) $GLOBALS["gTimer"] = new cTimer();

		// Open connection
		if (!isset($conn)) $conn = ew_Connect();
	}

	//
	//  Page_Init
	//
	function Page_Init() {
		global $gsExport, $gsExportFile, $UserProfile, $Language, $Security, $objForm;

		// Security
		$Security = new cAdvancedSecurity();
		$this->CurrentAction = (@$_GET["a"] <> "") ? $_GET["a"] :
@$_POST["a_list"]; // Set up curent action
```

```
		// Global Page Loading event (in userfn*.php)
		Page_Loading();

		// Page Load event
		$this->Page_Load();
	}

	//
	// Page_Terminate
	//
	function Page_Terminate($url = "") {
		global $conn;

		// Page Unload event
		$this->Page_Unload();

		// Global Page Unloaded event (in userfn*.php)
		Page_Unloaded();
		$this->Page_Redirecting($url);

		 // Close connection
		$conn->Close();
```

```
		// Go to URL if specified
		if ($url <> "") {
			if (!EW_DEBUG_ENABLED && ob_get_length())
				ob_end_clean();
			header("Location: " . $url);
		}
		exit();
	}
	var $Username;
	var $LoginType;

	//
	// Page main
	//
	function Page_Main() {
		global $Security, $Language, $UserProfile, $gsFormError;
		global $Breadcrumb;
		$Breadcrumb = new cBreadcrumb;
		$Breadcrumb->Add("login", "<span id=\"ewPageCaption\">" . $Language-
>Phrase("LoginPage") . "</span>", ew_CurrentUrl());
		$sPassword = "";
		$sLastUrl = $Security->LastUrl(); // Get last URL
		if ($sLastUrl == "")
```

```
				$sLastUrl = "index.php";
			if (IsLoggingIn()) {
				$this->Username =
@$_SESSION[EW_SESSION_USER_PROFILE_USER_NAME];
				$sPassword =
@$_SESSION[EW_SESSION_USER_PROFILE_PASSWORD];
				$this->LoginType =
@$_SESSION[EW_SESSION_USER_PROFILE_LOGIN_TYPE];
				$bValidPwd = $Security->ValidateUser($this->Username, $sPassword,
FALSE);
				if ($bValidPwd) {
					$_SESSION[EW_SESSION_USER_PROFILE_USER_NAME] =
"";
					$_SESSION[EW_SESSION_USER_PROFILE_PASSWORD] =
"";
					$_SESSION[EW_SESSION_USER_PROFILE_LOGIN_TYPE] =
"";
				}
			} else {
				if (!$Security->IsLoggedIn())
					$Security->AutoLogin();
				$this->Username = ""; // Initialize
				if (@$_POST["username"] <> "") {
```

```
// Setup variables
$this->Username =
ew_RemoveXSS(ew_StripSlashes(@$_POST["username"]));
$sPassword =
ew_RemoveXSS(ew_StripSlashes(@$_POST["password"]));
$this->LoginType =
strtolower(ew_RemoveXSS(@$_POST["type"]));
}
if ($this->Username <> "") {
$bValidate = $this->ValidateForm($this->Username, $sPassword);
if (!$bValidate)
$this->setFailureMessage($gsFormError);
$_SESSION[EW_SESSION_USER_PROFILE_USER_NAME] =
$this->Username; // Save login user name
$_SESSION[EW_SESSION_USER_PROFILE_LOGIN_TYPE] =
$this->LoginType; // Save login type
} else {
if ($Security->IsLoggedIn()) {
if ($this->getFailureMessage() == "")
$this->Page_Terminate($sLastUrl); // Return to last
accessed page
}
```

```
$bValidate = FALSE;

// Restore settings
if (@$_COOKIE[EW_PROJECT_NAME]['Checksum'] ==
strval(crc32(md5(EW_RANDOM_KEY))))
    $this->Username =
ew_Decrypt(@$_COOKIE[EW_PROJECT_NAME]['Username']);
if (@$_COOKIE[EW_PROJECT_NAME]['AutoLogin'] ==
"autologin") {
    $this->LoginType = "a";
} elseif (@$_COOKIE[EW_PROJECT_NAME]['AutoLogin'] ==
"rememberusername") {
    $this->LoginType = "u";
} else {
    $this->LoginType = "";
}
}
$bValidPwd = FALSE;
if ($bValidate) {

// Call Logging In event
$bValidate = $this->User_LoggingIn($this->Username,
$sPassword);
```

```
				if ($bValidate) {
					$bValidPwd = $Security->ValidateUser($this->Username,
$sPassword, FALSE); // Manual login
					if (!$bValidPwd) {
						if ($this->getFailureMessage() == "")
							$this->setFailureMessage($Language-
>Phrase("InvalidUidPwd")); // Invalid user id/password
					}
				} else {
					if ($this->getFailureMessage() == "")
						$this->setFailureMessage($Language-
>Phrase("LoginCancelled")); // Login cancelled
				}
			}
		}
		if ($bValidPwd) {

			// Write cookies
			if ($this->LoginType == "a") { // Auto login
				setcookie(EW_PROJECT_NAME . '[AutoLogin]',  "autologin",
EW_COOKIE_EXPIRY_TIME); // Set autologin cookie
				setcookie(EW_PROJECT_NAME . '[Username]',
ew_Encrypt($this->Username), EW_COOKIE_EXPIRY_TIME); // Set user name cookie
```

```
				setcookie(EW_PROJECT_NAME . '[Password]',
ew_Encrypt($sPassword), EW_COOKIE_EXPIRY_TIME); // Set password cookie
				setcookie(EW_PROJECT_NAME . '[Checksum]',
crc32(md5(EW_RANDOM_KEY)), EW_COOKIE_EXPIRY_TIME);
			} elseif ($this->LoginType == "u") { // Remember user name
				setcookie(EW_PROJECT_NAME . '[AutoLogin]',
"rememberusername", EW_COOKIE_EXPIRY_TIME); // Set remember user name cookie
				setcookie(EW_PROJECT_NAME . '[Username]',
ew_Encrypt($this->Username), EW_COOKIE_EXPIRY_TIME); // Set user name cookie
				setcookie(EW_PROJECT_NAME . '[Checksum]',
crc32(md5(EW_RANDOM_KEY)), EW_COOKIE_EXPIRY_TIME);
			} else {
				setcookie(EW_PROJECT_NAME . '[AutoLogin]', "",
EW_COOKIE_EXPIRY_TIME); // Clear auto login cookie
			}

			// Call loggedin event
			$this->User_LoggedIn($this->Username);
			$this->Page_Terminate($sLastUrl); // Return to last accessed URL
		} elseif ($this->Username <> "" && $sPassword <> "") {

			// Call user login error event
			$this->User_LoginError($this->Username, $sPassword);
```

```
	}
}

//
// Validate form
//
function ValidateForm($usr, $pwd) {
	global $Language, $gsFormError;

	// Initialize form error message
	$gsFormError = "";

	// Check if validation required
	if (!EW_SERVER_VALIDATE)
		return TRUE;
	if (trim($usr) == "") {
		ew_AddMessage($gsFormError, $Language->Phrase("EnterUid"));
	}
	if (trim($pwd) == "") {
		ew_AddMessage($gsFormError, $Language->Phrase("EnterPwd"));
	}

	// Return validate result
```

```
		$ValidateForm = ($gsFormError == "");

		// Call Form Custom Validate event
		$sFormCustomError = "";
		$ValidateForm = $ValidateForm && $this-
>Form_CustomValidate($sFormCustomError);
		if ($sFormCustomError <> "") {
			ew_AddMessage($gsFormError, $sFormCustomError);
		}
		return $ValidateForm;
	}

	// Page Load event
	function Page_Load() {

		//echo "Page Load";
	}

	// Page Unload event
	function Page_Unload() {

		//echo "Page Unload";
	}
```

```
}

// Page Redirecting event
function Page_Redirecting(&$url) {

	// Example:
	//$url = "your URL";

}

// Message Showing event
// $type = ''|'success'|'failure'
function Message_Showing(&$msg, $type) {

	// Example:
	//if ($type == 'success') $msg = "your success message";

}

// Page Render event
function Page_Render() {

	//echo "Page Render";
}
```

```
// Page Data Rendering event
function Page_DataRendering(&$header) {

	// Example:
	//$header = "your header";

}

// Page Data Rendered event
function Page_DataRendered(&$footer) {

	// Example:
	//$footer = "your footer";

}

// User Logging In event
function User_LoggingIn($usr, &$pwd) {

	// Enter your code here
	// To cancel, set return value to FALSE
```

```
		return TRUE;
	}

	// User Logged In event
	function User_LoggedIn($usr) {

		//echo "User Logged In";
	}

	// User Login Error event
	function User_LoginError($usr, $pwd) {

		//echo "User Login Error";
	}

	// Form Custom Validate event
	function Form_CustomValidate(&$CustomError) {

		// Return error message in CustomError
		return TRUE;
	}
}
?>
```

```
<?php ew_Header(FALSE) ?>
<?php

// Create page object
if (!isset($login)) $login = new clogin();

// Page init
$login->Page_Init();

// Page main
$login->Page_Main();

// Global Page Rendering event (in userfn*.php)
Page_Rendering();

// Page Rendering event
$login->Page_Render();
?>
<?php include_once "header.php" ?>
<script type="text/javascript">

// Write your client script here, no need to add script tags.
</script>
```

```
<script type="text/javascript">
var flogin = new ew_Form("flogin");

// Validate function
flogin.Validate = function()
{
	var fobj = this.Form;
	if (!this.ValidateRequired)
		return true; // Ignore validation
	if (!ew_HasValue(fobj.username))
		return this.OnError(fobj.username, ewLanguage.Phrase("EnterUid"));
	if (!ew_HasValue(fobj.password))
		return this.OnError(fobj.password, ewLanguage.Phrase("EnterPwd"));

	// Call Form Custom Validate event
	if (!this.Form_CustomValidate(fobj)) return false;
	return true;
}

// Form_CustomValidate function
flogin.Form_CustomValidate =
 function(fobj) { // DO NOT CHANGE THIS LINE!
```

```
        // Your custom validation code here, return false if invalid.
        return true;
 }

// Requires js validation
<?php if (EW_CLIENT_VALIDATE) { ?>
flogin.ValidateRequired = true;
<?php } else { ?>
flogin.ValidateRequired = false;
<?php } ?>
</script>
<?php $Breadcrumb->Render(); ?>
<?php $login->ShowPageHeader(); ?>
<?php
$login->ShowMessage();
?>
<form name="flogin" id="flogin" class="ewForm form-horizontal" action="<?php echo
ew_CurrentPage() ?>" method="post">
<div class="ewLoginContent">
        <div class="control-group">
                <label class="control-label" for="username"><?php echo $Language-
>Phrase("Username") ?></label>
```

```
<div class="controls"><input type="text" name="username" id="username"
class="input-large" value="<?php echo $login->Username ?>" placeholder="<?php echo
$Language->Phrase("Username") ?>"></div>
        </div>
        <div class="control-group">
                <label class="control-label" for="password"><?php echo $Language-
>Phrase("Password") ?></label>
                <div class="controls"><input type="password" name="password" id="password"
class="input-large" placeholder="<?php echo $Language->Phrase("Password") ?>"></div>
        </div>
        <div class="control-group">
                <div class="controls">
                <label class="radio ewRadio" style="white-space: nowrap;"><input type="radio"
name="type" id="type" value="a"<?php if ($login->LoginType == "a") { ?>
checked="checked"<?php } ?>><?php echo $Language->Phrase("AutoLogin") ?></label>
                <label class="radio ewRadio" style="white-space: nowrap;"><input type="radio"
name="type" id="type" value="u"<?php if ($login->LoginType == "u") { ?>
checked="checked"<?php } ?>><?php echo $Language->Phrase("SaveUserName") ?></label>
                <label class="radio ewRadio" style="white-space: nowrap;"><input type="radio"
name="type" id="type" value=""<?php if ($login->LoginType == "") { ?>
checked="checked"<?php } ?>><?php echo $Language->Phrase("AlwaysAsk") ?></label>
                </div>
        </div>
```

```
	<div class="control-group">
		<div class="controls">
			<button class="btn btn-primary ewButton" name="btnsubmit"
id="btnsubmit" type="submit"><?php echo $Language->Phrase("Login") ?></button>
		</div>
	</div>
</div>
</form>
<br>
<script type="text/javascript">
flogin.Init();
<?php if (EW_MOBILE_REFLOW && ew_IsMobile()) { ?>
ew_Reflow();
<?php } ?>
</script>
<?php
$login->ShowPageFooter();
if (EW_DEBUG_ENABLED)
	echo ew_DebugMsg();
?>
<script type="text/javascript">

// Write your startup script here
```

```
// document.write("page loaded");


</script>
<?php include_once "footer.php" ?>
<?php
$login->Page_Terminate();
?>


//message
<?php require_once "core/controller.php";?>
<?php
if(!isset($_SESSION['user_id']))
    echo "Not Logged In";
?>
<?php


$id = $_GET['mid'];


$sql = mysqli_query($con, "SELECT * FROM messages WHERE id = '$id' ");


$dta = mysqli_fetch_object($sql);
```

```
$sql1 = mysqli_query($con, "UPDATE messages SET status = 1 WHERE id = '$id' ");


?>

<!DOCTYPE html>
<html lang="en">

<head>
        <title><?php echo $_SESSION['name'] ?></title>
        <?php include "inc_head.php";?>

</head>
    <body data-color="grey" class="flat">
        <div id="wrapper">
            <div id="header">
                <h1><a href="index-2.html">Unicorn Admin</a></h1>
                <a id="menu-trigger" href="#"><i class="fa fa-bars"></i></a>
            </div>


            <?php include "inc_menu.php";?>
```

```
        <div id="content">
            <div id="content-header" class="mini">
                <h1><?php echo $_SESSION['name'] ?></h1>
            </div>

            <div id="breadcrumb">
                <a href="#" title="Go to Home" class="tip-bottom"><i class="fa fa-home"></i>
Home</a>
                <a href="#" class="current">Dashboard</a>
            </div>

            <div class="container">
                <div class="widget-box">
                        <div class="widget-title">
                            <span class="icon">
                                <i class="fa fa-bell"></i>
                            </span>
                            <h5>Message Details</h5>
                        </div>
                        <div class="widget-content">
                        <div class="row">
                            <div class="col-md-12">
```

```
                              <p>
                                 <?php echo $dta->message ?>
                              </p>
                              <small>Sent: <?php echo date("D M j, H:i", strtotime($dta->created_at))
?></small>
                           </div>
                        </div>
                        </div>
                     </div>
                  </div>

               </div>
               <?php include "inc_footer.php";?>
            </div>

               <?php include "inc_scripts.php";?>

         </body>

</html>
```

```
//school
<?php require_once "core/controller.php";?>
<?php
if(!isset($_SESSION['user_id']))
    echo "Not Logged In";

if($_SESSION['type'] != 'S')
    header('Location: logout.php')
?>
<?php

$sql = mysqli_query($con, "SELECT * FROM users WHERE status = 'Verified' AND type = 'C'
");



?>


<!DOCTYPE html>
<html lang="en">


<head>
        <title>Dashboard</title>
        <?php include "inc_head.php";?>
```

```
</head>
    <body data-color="grey" class="flat">
        <div id="wrapper">
            <div id="header">
                <h1><a href="index-2.html">Unicorn Admin</a></h1>
                <a id="menu-trigger" href="#"><i class="fa fa-bars"></i></a>
            </div>



            <?php include "inc_menu.php";?>

            <div id="content">
                <div id="content-header" class="mini">
                    <h1><?php echo $_SESSION['name'] ?></h1>
                </div>

                <div id="breadcrumb">
                    <a href="#" title="Go to Home" class="tip-bottom"><i class="fa fa-home"></i>
Home</a>
                    <a href="#" class="current">Dashboard</a>
                </div>
```

```
<div class="container">
    <div class="widget-box">
        <div class="widget-title">
            <span class="icon">
                <i class="fa fa-bell"></i>
            </span>
            <h5>Send Message</h5>
        </div>
        <div class="widget-content">
        <div class="row">
            <div class="col-md-12">
                <form id="msg-form" action="core/controller.php" method="post">
                  <!-- <div class="form-group">
                    <label for="exampleInputEmail1">Student Name</label>
                    <input type="text" name="sname" class="form-control"
placeholder="Student Name">
                  </div> -->
                  <div class="form-group">
                    <label for="">Select Company</label>
                    <select name="school" class="form-control">
                        <option value="">---Select---</option>
                        <?php while ($data = mysqli_fetch_object($sql)) { ?>
```

```
                                        <option value="<?php echo $data->id?>"><?php echo $data->name?></option>
                                    <?php } ?>
                                </select>
                            </div>

                            <div class="form-group">
                                <label> Message: </label>
                                <textarea name="message" class="form-control" id="" cols="30" rows="10"></textarea>
                            </div>

                            <button type="submit" class="btn btn-default">Submit</button>
                        </form>
                    </div>
                </div>
                </div>
            </div>
        </div>


    </div>
    <?php include "inc_footer.php";?>
</div>
```

```
        <?php include "inc_scripts.php";?>
<script>
  $('#msg-form').on('submit', function(){
    var that = $(this),
        url  = that.attr('action'),
        method = that.attr('method'),
        data = {};

        that.find('[name]').each(function (index, value) {
            var that = $(this),
            name = that.attr('name'),
            value = that.val();

            data[name] = value;
        });
        // console.log(data);

        $.ajax({
            url: url,
            type: method,
            data: data,
            success: function(response){
```

```
                    console.log(response);
                    $('#msg-form').trigger('reset')
                    alert("Message Sent")
                }
            })
        return false;
    })
</script>
    </body>

</html>


//logout
<?php
session_start();
session_destroy();
header('location:index.php');

?>
```